\documentclass[aps, pra, reprint, groupedaddress, twoside]{revtex4-2}

\usepackage{graphicx}
\usepackage{amsfonts}
\usepackage{amsmath}
\usepackage{amssymb}
\usepackage{enumerate}

\usepackage{xcolor}
\definecolor{grey}{rgb}{0.7,0.7,0.7}
\definecolor{db}{rgb}{0,0,0.5}

\usepackage{fancyhdr}
 
\begin{document}

\title{\colorbox{db}{\parbox{\linewidth}{  \centering \parbox{0.9\linewidth}{ \textbf{\centering\Large{\color{white}{ \vskip0.2em{ Generalized Matrix Transformation Formalism for Reflection and Transmission of Complex Optical Waves at a Plane Dielectric Interface}\vskip0.8em}}}}}}}


\author{Anirban Debnath}
\email[]{anirban.debnath090@gmail.com}
\affiliation{School of Physics, University of Hyderabad, Hyderabad --- 500046, India}

\author{Nirmal K. Viswanathan}
\email[]{nirmalsp@uohyd.ac.in}
\affiliation{School of Physics, University of Hyderabad, Hyderabad --- 500046, India}

\date{\today}

\begin{abstract}
\noindent{\color{grey}{\rule{0.784\textwidth}{1pt}}}
\vspace{-0.8em}

We describe a generalized formalism, addressing the fundamental problem of reflection and transmission of complex optical waves at a plane dielectric interface. Our formalism involves the application of generalized operator matrices to the incident constituent plane wave fields to obtain the reflected and transmitted constituent plane wave fields. We derive these matrices and describe the complete formalism by implementing these matrices. This formalism, though physically equivalent to Fresnel formalism, has greater mathematical elegance and computational efficiency as compared to the latter. 
We utilize exact 3D expressions of the constituent plane wavevectors and electric fields of the incident, reflected and transmitted waves, which enable us to seamlessly analyse plane waves, paraxial and non-paraxial beams, highly diverging and tightly focused beam-fields as well as waves of miscellaneous wavefront-shapes and properties using the single formalism. 
The exact electric field expressions automatically include the geometric phase information; while we retain the wavefront curvature information by using appropriate multiplicative factors.
We demonstrate our formalism by obtaining the reflected and transmitted fields in a simulated Gaussian beam model.
Finally, we briefly discuss how our generalized formalism is capable of analysing the reflection-transmission problem of a very large class of complex optical waves --- 
by referring to some novel works from the current literature as exemplary cases.
{\color{grey}{\rule{0.784\textwidth}{1pt}}}
\end{abstract}

\maketitle


\tableofcontents

\newpage

\section{Introduction}

The reflection and transmission of a plane electromagnetic wave at a plane isotropic dielectric interface (`isotropic' is omitted here onwards) is a well-studied problem of fundamental interest in electromagnetic theory. The problem refers to finding the reflected and transmitted wavevectors and electric (and magnetic) fields for a given pair of incident wavevector and electric (and magnetic) field. The standard solution to this problem, as worked out in textbooks on electromagnetic theory and optics \cite{Jackson, SalehTeich, BornWolf},  determines the reflected and transmitted wavevectors by using phase matching. These results are also re-expressed as the geometrical laws of reflection and transmission. The reflected and transmitted electric fields are then determined by using electromagnetic boundary conditions at the dielectric interface. As the final result, the reflected and transmitted electric field amplitude values for the transverse magnetic (TM) and transverse electric (TE) polarizations are expressed as scalar multiples of the incident electric field amplitude values of the corresponding polarizations. The multiplicative factors are the Fresnel coefficients.

However, no physical wave is an ideal plane wave. Diverging and converging beams (both paraxial and non-paraxial) \cite{SalehTeich}, spherical and cylindrical waves \cite{SalehTeich}, vectorial vortex beams with helical wavefronts \cite{Gbur, Poynting, RABeth, LAllen, HHe, PA2000, Yao, DennisVortex, BNRev, UriLevyRev} 
are some of the common optical waves used in different optical experiments. The current literature is rich with novel research works describing various effects arising due to the reflection and transmission of complex optical waves at plane dielectric interfaces. Goos-H\"anchen (GH) and Imbert-Fedorov (IF) shifts of paraxial beams in this context have been well studied by many researchers \cite{GH, Artmann, RaJW, AntarYM, McGuirk, ChanCC, Porras, AielloArXiv, Fedorov, Schilling, Imbert, Player, FVG, Liberman, Onoda, Bliokh2006, Bliokh2007, HostenKwiat, AielloArXiv2, Aiello2008, Aiello2009, Qin2011, BARev}. Bliokh and Aiello \cite{BARev} have described the reflection and transmission of paraxial beams in terms of effective Jones matrices, using which they have described a generalized theory of GH and IF shifts of such beams. Dennis, G\"otte and L\"offler \cite{Dennis, Gotte, GotteLofflerDennis} have studied the field-properties of GH and IF shifted beams; and have established their connection to quantum mechanical weak measurement \cite{AAV, DSS, RSH, Dennis, Gotte, HostenKwiat, Aiello2008}. Berry \cite{Berry435} has carried out an exact analysis of reflected dipole radiation field --- an example of a highly diverging optical wave-field. Spin-orbit interaction (SOI) and spin Hall effect of light (SHEL), arising due to the reflection and transmission of optical beams, have been observed by many researchers \cite{Liberman, Onoda, Bliokh2006, Bliokh2007, HostenKwiat, Qin2011, XieSHELinIF}. 
Li et al. \cite{LiVortexRT}, Dennis-G\"otte \cite{DennisGotteVortex} and Yavorsky-Brasselet \cite{YavorskyBrasselet} have studied the reflection and transmission of vortex beams. Vectorial vortex beam generation due to the reflection and transmission of non-vortex beams have been observed by 
Barczyk et al. \cite{VortexBrewster}. Brekhovskikh \cite{LMB} has carried out an extensive analysis on reflection and transmission of waves in layered media.

The presence of all such effects has been brought to the fore by using the 
same underlying principle.
By using standard methods like Fourier decomposition, a complex wave is expressed as a combination of many constituent plane waves having different wavevectors and electric fields \cite{SalehTeich}. 
In the reflection-transmission scenario, each constituent plane wave is treated in the standard way by using Fresnel coefficients. After obtaining the reflected and transmitted plane waves for each incident constituent plane wave, all the corresponding output plane waves are combined together to obtain the complete reflected and transmitted waves.
Since the plane-of-incidence--angle-of-incidence pair is unique for each wavevector, the different constituent plane waves reflect/transmit differently due to the differences in Fresnel coefficients. As a result, we obtain unique electric field profiles of the complete reflected and transmitted waves. These unique profiles, which are generally considerably different from the complete incident field profile, are the keys to all the novel effects mentioned above.

Though most widely used, Fourier decomposition is not the only way for expressing an optical wave in terms of a combination of many plane waves. 
In some cases, the given wavefront can be divided into surface elements, which locally behave as plane waves. This decomposition is particularly convenient for waves with simple-shaped wavefronts; e.g. spherical and cylindrical waves. When such a wave is incident on a plane dielectric interface, Fresnel coefficients can be applied to the TM-TE electric field components of each surface element in the same way as are applied to the Fourier-decomposed plane-wave field components. The collection of all the reflected and transmitted surface elements then produce the respective complete waves. Thus, the application of Fresnel coefficients in finding the reflected and transmitted constituent plane-wave fields is a general step irrespective of the considered plane-wave decomposition method.


The application of Fresnel coefficients is a complete and self-consistent formalism from a physical perspective. However, from a mathematical perspective, it lacks algorithmic straightforwardness. 
The Fresnel coefficients are `relative amplitude' values \cite{Jackson} of the TM and TE components of the reflected and transmitted fields with respect to the amplitude values of the corresponding incident field components. 
But we observe a sign-ambiguity in this representation regarding the choice of coordinate systems.
Most authors (e.g. Refs. \cite{Jackson, BornWolf}) derive the Fresnel coefficient expressions by choosing the vector directions in a specific way. These expressions are applicable to the choice of beam coordinate systems as in Ref. \cite{BARev}. However, Ref. \cite{SalehTeich} uses a different choice of beam coordinate systems and obtains a different expression for the Fresnel TM reflection coefficient, which is negative of the corresponding expressions derived in Refs. \cite{Jackson, BornWolf}. Thus, it is essential to specify the coordinate systems and/or vector directions, with respect to which the Fresnel coefficients are determined.

Also, the TM-TE decomposition is carried out with respect to the plane of incidence. So, for a composite beam, we must transform the electric field corresponding to each constituent wavevector to the local coordinate system of that wavevector, so that the TM-TE decomposition can be carried out (e.g. Ref. \cite{Berry435}). Though Bliokh-Aiello have simplified the problem for paraxial beams \cite{BARev}, their formalism is not applicable to arbitrary complex waves.

So, after plane-wave decomposition of the incident wave and before recombination of the reflected and transmitted plane waves, the regular way of using the Fresnel coefficients includes several intermediate calculating steps for each pair of incident constituent wavevector and electric field.
This makes the calculating algorithm cumbersome; and also makes it computationally inefficient while simulating a model for reflection and transmission of complex optical waves.

We envision that the above difficulties can be avoided by abandoning the `relative amplitude' representation altogether. Instead, in the intermediate steps, the incident, reflected and transmitted electric fields must always 
be expressed in their full vector forms in terms of a single coordinate system --- the global dielectric-interface coordinate system. The nature of the reflection-transmission problem suggests that the reflected and transmitted plane-wave fields can be expressed as certain transforms of the incident plane-wave field. The transformation operators can be expressed in matrix forms in the same dielectric-interface coordinate system. We refer to these matrices as the reflection and transmission coefficient matrices. With these matrices, we can 
replace all the usual intermediate steps by essentially two new intermediate steps: (i) determination of the matrix elements; and (ii) application of the matrices to the incident constituent plane-wave fields to determine the corresponding reflected and transmitted plane-wave fields.
Thus, though this algorithm is physically equivalent to the usual Fresnel coefficient approach, it is distinguished by its mathematical elegance and computational efficiency. 
This algorithm is the content of our generalized matrix transformation formalism, the potential of which we demonstrate in the present work and in subsequent related works.

At all stages, we use exact 3D expressions of each constituent plane wavevector and electric field of the incident, reflected and transmitted waves, without making any special approximation. This enables us to analyse the reflection-transmission problems of a large class of complex optical waves of arbitrary divergence. We retain the information of wavefront curvature by using appropriate multiplying factors; and the information on the geometric phase \cite{P1956, Berry, Berry1987, Shapere, Bliokh2008, Bliokh2009, BA2010, BARev} is automatically retained by the exact expressions of the constituent plane-wave electric fields.

In this paper, we first derive the above-mentioned reflection and transmission coefficient matrices; and discuss their relation to Fresnel coefficients [Section \ref{Sec_Derivation}]. We then describe the complete mathematical formalism, by which these matrices are implemented in an actual reflection-transmission problem [Section \ref{Sec_Formalism}]. 
We then demonstrate our formalism by generating and analysing computational data in a simulated Gaussian beam model [Section \ref{Sec_Simulation}].
Then we give examples of some direct applications of our formalism by referring to novel optical phenomena from the current literature; 
and briefly discuss how to further generalize our formalism for total internal reflection problems [Section \ref{Sec_ApplGen}].
Finally, we briefly discuss the gain in computational efficiency and numerical accuracy achieved by the implementation of our formalism [Section \ref{Sec_CompAspects}].


\section{Notations and Conventions} \label{Context}

The notations and conventions which we use are as follows:

\begin{enumerate}

\item[\textbf{1.}] The media of incidence/reflection and transmission have refractive indices $n_1$ and $n_2$ respectively.
 
\item[\textbf{2.}] The functional forms of the wavevectors and electric fields of the complete waves are denoted by $\tilde{\mathbf{k}}_j$ and $\tilde{\mathbf{E}}_j$, where $j = i,r,t$ denote incidence, reflection and transmission. The field amplitude vector functions, phase functions and intensities of $\tilde{\mathbf{E}}_j$ are denoted by $\tilde{\boldsymbol{ \mathcal{E}}}_j$, $\tilde{\Phi}_j$ and $\mathcal{I}_j$ respectively. Our electric field amplitude and phase conventions are given in Appendix \ref{App_Amplitude}.

\item[\textbf{3.}] The constituent plane wavevectors and electric fields corresponding to $\tilde{\mathbf{k}}_j$ and $\tilde{\mathbf{E}}_j$ are denoted by $\mathbf{k}_j$ and $\mathbf{E}_j$ respectively, with the field amplitude vector being $\boldsymbol{ \mathcal{E}}_j$.

\item[\textbf{4.}] The central axial directions $\mathbf{k}_{j0}$ are defined as per geometrical convenience. 
For example, for the case of partial reflection and transmission of an optical beam, these are the central wavevectors and are also related via the geometrical laws of reflection and transmission. For dipole radiation or any other spherical wave, $\mathbf{k}_{j0}$ can be defined with respect to the perpendicular line joining the point source to the dielectric interface. 
The $\mathbf{k}_{j0}$ directions are chosen only to define the global beam coordinate systems; and the main electromagnetic calculations are not affected by these choices.


\item[\textbf{5.}] The global dielectric-interface coordinate system $S$ and the global beam coordinate systems $J=I,R,T$ are defined with respect to the central axial directions $\mathbf{k}_{j0}$ in the way as shown in Fig. \ref{fig_co}. The main electromagnetic calculations are carried out in the $S$ coordinate system. The beam coordinate systems are used to define the incident electric field (the $I$ coordinate system) and to observe the reflected and transmitted electric fields (the $J' = R,T$ coordinate systems).

\begin{figure}[t]
\centering
\includegraphics[width = 0.58\linewidth]{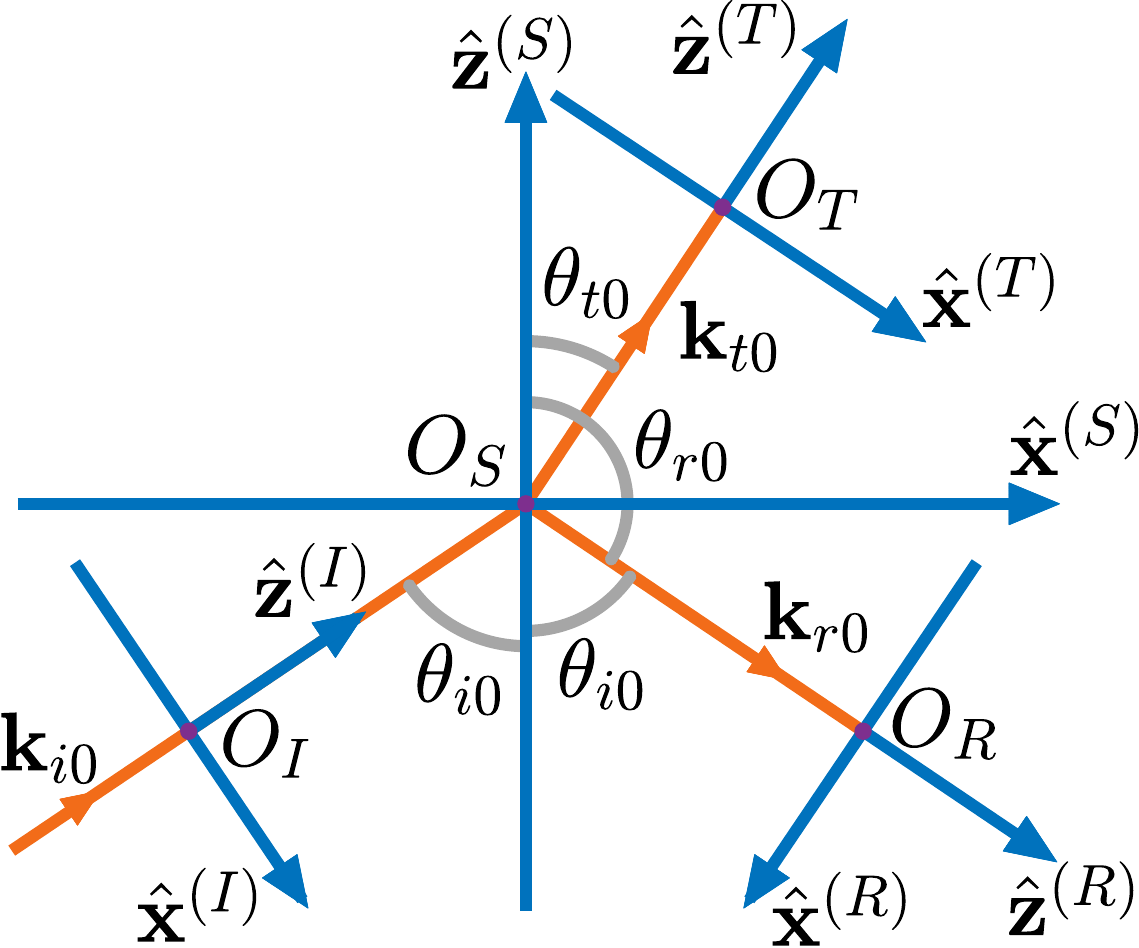}
\caption{
\small{The global dielectric-interface coordinate system $S$ and the global beam coordinate systems $J = I, R, T$; defined with respect to the central directions $\mathbf{k}_{j0}$.}
}
{\color{grey}{\rule{\linewidth}{1pt}}}
\label{fig_co}
\end{figure}

\item[\textbf{6.}] Any vector $\mathbf{V}_j$ in a coordinate system $C$ are expressed as $\mathbf{V}_j^{(C)} = V_{jx}^{(C)} \hat{\mathbf{x}}^{(C)} + V_{jy}^{(C)} \hat{\mathbf{y}}^{(C)} + V_{jz}^{(C)} \hat{\mathbf{z}}^{(C)}$; where, the superscript `$(C)$' is the coordinate system identifier.

\item[\textbf{7.}] The $\mathbf{k}_{j0}$ directions make angles $\theta_{j0}$ with $\hat{\mathbf{z}}^{(S)}$. In particular, if $\mathbf{k}_{j0}$ are central wavevetors related via the geometrical laws of reflection and transmission, then $\theta_{r0} = \pi - \theta_{i0}$, $\theta_{t0} = \sin^{-1} \left(n_1 \sin\theta_{i0}/n_2\right)$.

\item[\textbf{8.}] For any coordinate system $C$, the $z^{(C)}=0$ plane is referred to as the surface $C$; and any quantity $\mathbf{Q}_j$ on that surface is denoted by $\mathbf{Q}_{jC}$. Thus, an expression of the form $\mathbf{Q}_{jC_1}^{(C_2)}$ refers to the quantity $\mathbf{Q}_{j}$ considered at the surface $C_1$ and expressed in terms of the coordinate system $C_2$. Also, wherever relevant, we denote the $z^{(C)}$ coordinate of a point at the surface $C$ as $0^{(C)}$.

\item[\textbf{9.}] The transformations of wavevectors and fields among different coordinate systems are written in the forms $\mathbf{V}_j^{(S)} = \tilde{\mathbf{R}}_{SJ} \mathbf{V}_j^{(J)}$ and $\mathbf{V}_j^{(J)} = \tilde{\mathbf{R}}_{JS} \mathbf{V}_j^{(S)}$, where $\tilde{\mathbf{R}}_{JS} = \tilde{\mathbf{R}}_{SJ}^{-1}$ are standard rotation matrices.

\end{enumerate}


\section{The Reflection and Transmission Coefficient Matrices}\label{Sec_Derivation}

Using the above notations and conventions, we derive the reflection and transmission coefficient matrices at surface $S$ by considering only one incident constituent plane wavevector $\mathbf{k}_{iS}^{(S)}$, and its corresponding electric field $\mathbf{E}_{iS}^{(S)}$. We perform the calculations by using vector forms of $\mathbf{k}_{iS}^{(S)}$ and $\mathbf{E}_{iS}^{(S)}$ at all stages, instead of using their amplitude values. We 
derive the transformation matrices using electromagnetic boundary conditions; and later establish their relation to Fresnel coefficients.

\subsection{Derivation}

The fields at the surface $S$ satisfy the phase matching condition \cite{SalehTeich, BornWolf, Jackson}, using which we obtain
\begin{eqnarray}
&\mathbf{k}_{j'S}^{(S)} = \tilde{\mathbf{I}}_{j'S}^{(S)}\,\mathbf{k}_{iS}^{(S)}; &\label{kjSS}\\ 
&\mbox{where,}\hspace{1em} \tilde{\mathbf{I}}_{j'S}^{(S)} = \left[\begin{array}{ccc}
1 & 0 & 0 \\
0 & 1 & 0 \\
0 & 0 & k_{j'zS}^{(S)}/k_{izS}^{(S)}
\end{array}\right] \! ;
\label{Imat}&
\\
& k_{rzS}^{(S)} = -k_{izS}^{(S)};
\hspace{1em} 
k_{tzS}^{(S)} = \left[ n_2^2 k^2 - \left( n_1^2 k^2 - k_{izS}^{(S)\;2} \right) \right]^\frac{1}{2} \!\! .\hspace{0.8em}&\label{kjzSS}
\end{eqnarray}
These results can also be expressed as the geometrical laws of reflection and transmission. 

Once the phase matching is established, we carry out the rest of the calculations only in terms of the electric field amplitude vectors $\boldsymbol{\mathcal{E}}_{jS}^{(S)}$. The components of these vectors can be complex, depending upon the ellipticity of polarization [Appendix \ref{App_Amplitude}]. The corresponding magnetic field amplitude vectors are [Appendix \ref{App_CompAmpConditions}]
\begin{equation}
\boldsymbol{\mathcal{H}}_{jS}^{(S)} = \left( \mathbf{k}_{jS}^{(S)} \times \boldsymbol{\mathcal{E}}_{jS}^{(S)} \right)/\omega\mu_0.\label{H=kcrossE}
\end{equation}
The electromagnetic boundary conditions are satisfied by $\boldsymbol{\mathcal{E}}_{jS}^{(S)}$ and $\boldsymbol{\mathcal{H}}_{jS}^{(S)}$ [Appendix \ref{App_CompAmpConditions}]: 
\begin{eqnarray}
&\mathcal{E}_{ixS}^{(S)} + \mathcal{E}_{rxS}^{(S)} = \mathcal{E}_{txS}^{(S)};& \label{Exbc}\\
&\mathcal{E}_{iyS}^{(S)} + \mathcal{E}_{ryS}^{(S)} = \mathcal{E}_{tyS}^{(S)};&\label{Eybc}\\
&\mathcal{H}_{ixS}^{(S)} + \mathcal{H}_{rxS}^{(S)} = \mathcal{H}_{txS}^{(S)};&\label{Hxbc}\\
&\mathcal{H}_{iyS}^{(S)} + \mathcal{H}_{ryS}^{(S)} = \mathcal{H}_{tyS}^{(S)}.\label{Hybc}&
\end{eqnarray}
Also, since the dielectric media are isotropic, $\mathbf{k}_{jS}^{(S)}$ and $\boldsymbol{\mathcal{E}}_{jS}^{(S)}$ satisfy orthogonality [Appendix \ref{App_CompAmpConditions}]: 
\begin{equation}
\mathbf{k}_{jS}^{(S)} \cdot \boldsymbol{\mathcal{E}}_{jS}^{(S)} = 0. \label{k.E=0}
\end{equation}
Solving Eqs. (\ref{Exbc}--\ref{k.E=0}) simultaneously [Appendix \ref{App_CentralResult}], 
we obtain the reflected and transmitted field amplitude vectors as
\begin{eqnarray}
&\boldsymbol{\mathcal{E}}_{j'S}^{(S)} = \tilde{\mathbf{j'}}_{S}^{(S)} \boldsymbol{\mathcal{E}}_{iS}^{(S)},\hspace{1em} (j' = r, t);&\label{EjSS}\\
&\mbox{where,}\hspace{1em}\tilde{\mathbf{r}}_{S}^{(S)} = A_0 
\begin{bmatrix}
A_{11} & A_{xy} & 0 \\
A_{xy} & -A_{10} & 0 \\
0 & 0 & -A_{01}
\end{bmatrix};&
\label{rMat}\\
& \tilde{\mathbf{t}}_{S}^{(S)} = 
\begin{bmatrix}
1 + A_0 A_{11} & A_0 A_{xy} & 0 \\
A_0 A_{xy} & 1 - A_0 A_{10} & 0 \\
0 & 0 & \dfrac{k_{izS}^{(S)}}{k_{tzS}^{(S)}} (1 + A_0 A_{01})
\end{bmatrix} \! ;
\hspace{1.5em}&
\label{tMat}
\end{eqnarray}
\begin{subequations}
\label{Aterms}
\begin{equation}
A_{pq} = k_{ixS}^{(S)\,2} + (-1)^p\, k_{iyS}^{(S)\,2} + (-1)^q\, k_{tzS}^{(S)}k_{izS}^{(S)},\hspace{0.5em}(p,q = 0,1);\label{Apq}
\end{equation}
\vspace{-1em}
\begin{equation}
\hspace{-0.4em}A_{xy} = 2\, k_{ixS}^{(S)} k_{iyS}^{(S)};
\hspace{1em}
A_z = \dfrac{k_{tzS}^{(S)} - k_{izS}^{(S)}}{k_{tzS}^{(S)} + k_{izS}^{(S)}}; \hspace{1em} A_0 = \dfrac{A_z}{A_{00}}.\label{Axyz0}
\end{equation}
\end{subequations}
The matrices $\tilde{\mathbf{j'}}_{S}^{(S)}$ are the reflection and transmission coefficient matrices. 
Thus, the reflection and transmission coefficient matrices are operators, which can transform a given incident plane-wave field amplitude vector $\boldsymbol{\mathcal{E}}_{iS}^{(S)}$ (or, the field $\mathbf{E}_{iS}^{(S)}$ in general) to the reflected and transmitted plane-wave field amplitude vectors $\boldsymbol{\mathcal{E}}_{j'S}^{(S)}$ (or, to the fields $\mathbf{E}_{j'S}^{(S)}$ in general).

\subsection{Alternative Derivation}\label{Sub_AltDerivation}

The matrices $\tilde{\mathbf{j'}}_S^{(S)}$ [Eqs. (\ref{rMat}, \ref{tMat})] can also be derived by performing calculations with reference to the plane of incidence.
The plane of incidence is defined as the plane that contains the wavevector $\mathbf{k}_{iS}^{(S)}$ and the surface-normal $\hat{\mathbf{z}}^{(S)}$. By Eq. (\ref{kjSS}), the wavevectors $\mathbf{k}_{j'S}^{(S)}$ are also contained in the same plane. We define a coordinate system $S'$ such that $\hat{\mathbf{z}}^{(S')} = \hat{\mathbf{z}}^{(S)}$ and
\begin{eqnarray}
& k_{ixS}^{(S')}\, \hat{\mathbf{x}}^{(S')} = k_{ixS}^{(S)}\, \hat{\mathbf{x}}^{(S)} + k_{iyS}^{(S)}\, \hat{\mathbf{y}}^{(S)};&\\
& k_{ixS}^{(S')} = \left( k_{ixS}^{(S)\, 2} + k_{iyS}^{(S)\, 2} \right)^\frac{1}{2}.&
\end{eqnarray}
Then, the plane of incidence is the $y^{(S')} = 0$ plane. The wavevectors and electric fields can be transformed from the $S$ coordinate system to the $S'$ coordinate system by applying the rotation matrix
\begin{equation}
\tilde{\mathbf{R}}_{S'S} = 
\begin{bmatrix}
k_{ixS}^{(S)}/k_{ixS}^{(S')} & k_{iyS}^{(S)}/k_{ixS}^{(S')} & 0 \\
-k_{iyS}^{(S)}/k_{ixS}^{(S')} & k_{ixS}^{(S)}/k_{ixS}^{(S')} & 0 \\
0 & 0 & 1
\end{bmatrix}.
\label{RSPS}
\end{equation}

Corresponding to the electric field amplitude vectors $\boldsymbol{\mathcal{E}}_{jS}^{(S')}$, the magnetic field amplitude vectors are given by 
\begin{equation}
\boldsymbol{\mathcal{H}}_{jS}^{(S')} = \left( \mathbf{k}_{jS}^{(S')} \times \boldsymbol{\mathcal{E}}_{jS}^{(S')} \right)/\omega\mu_0.\label{HSP=kcrossESP}
\end{equation}
The vectors $\mathbf{k}_{jS}^{(S')}$, $\boldsymbol{\mathcal{E}}_{jS}^{(S')}$ and $\boldsymbol{\mathcal{H}}_{jS}^{(S')}$ satisfy the same boundary and orthogonality conditions as Eqs. (\ref{Exbc}--\ref{k.E=0}), with the coordinate system $S$ now replaced by $S'$. 
Simultaneously solving the resulting equations, we obtain 
\begin{eqnarray}
&\boldsymbol{\mathcal{E}}_{j'S}^{(S')} = \tilde{\mathbf{j'}}_{S}^{(S')} \boldsymbol{\mathcal{E}}_{iS}^{(S')},\hspace{1em} (j' = r, t);&\label{EjSSP}\\
&\mbox{where,}\hspace{1em}\tilde{\mathbf{r}}_{S}^{(S')} = 
\begin{bmatrix}
A_0 A_{01} & 0 & 0 \\
0 & -A_{z} & 0 \\
0 & 0 & -A_0 A_{01}
\end{bmatrix};&
\label{rMatSP}\\
& \tilde{\mathbf{t}}_{S}^{(S')} = 
\begin{bmatrix}
1 + A_0 A_{01} & 0 & 0 \\
0 & 1 - A_z & 0 \\
0 & 0 & \dfrac{k_{izS}^{(S')}}{k_{tzS}^{(S')}} (1 + A_0 A_{01})
\end{bmatrix} \! ;
\hspace{1.5em}&
\label{tMatSP}
\end{eqnarray}
where, the terms $A_{pq}$ ($p,q = 0,1$), $A_z$, $A_0$ are defined by Eqs. (\ref{Aterms}). Then, transforming Eq. (\ref{EjSSP}) from the $S'$ coordinate system to the $S$ coordinate system, Eq. (\ref{EjSS}) is reproduced:
\begin{eqnarray}
\tilde{\mathbf{R}}_{SS'} \boldsymbol{\mathcal{E}}_{j'S}^{(S')} &=& \tilde{\mathbf{R}}_{SS'} \tilde{\mathbf{j'}}_{S}^{(S')}  \boldsymbol{\mathcal{E}}_{iS}^{(S')} \nonumber\\ 
&=& \tilde{\mathbf{R}}_{SS'} \tilde{\mathbf{j'}}_{S}^{(S')} \tilde{\mathbf{R}}_{S'S} \tilde{\mathbf{R}}_{SS'}  \boldsymbol{\mathcal{E}}_{iS}^{(S')}; \nonumber
\end{eqnarray}
\vspace{-1.5em}
\begin{eqnarray}
\mbox{or,} && \hspace{1em} \boldsymbol{\mathcal{E}}_{j'S}^{(S)} = \tilde{\mathbf{j'}}_{S}^{(S)} \boldsymbol{\mathcal{E}}_{iS}^{(S)};\hspace{2em}\mbox{[Eq. (\ref{EjSS})];} \nonumber\\
\mbox{where,} && \hspace{1.07em} \tilde{\mathbf{j'}}_{S}^{(S)} = \tilde{\mathbf{R}}_{SS'} \tilde{\mathbf{j'}}_{S}^{(S')} \tilde{\mathbf{R}}_{S'S}.\label{j=RjPR}
\end{eqnarray}
Using Eqs. (\ref{RSPS}, \ref{rMatSP}, \ref{tMatSP}) in Eq. (\ref{j=RjPR}), we reproduce Eqs. (\ref{rMat}, \ref{tMat}).

The forms of Eqs. (\ref{rMatSP}, \ref{tMatSP}) imply that the eigenvectors of the $\tilde{\mathbf{j'}}_{S}^{(S)}$ (or $\tilde{\mathbf{j'}}_{S}^{(S')}$) matrices are $\hat{\mathbf{x}}^{(S')}$, $\hat{\mathbf{y}}^{(S')}$, $\hat{\mathbf{z}}^{(S')}$. Corresponding to these eigenvectors, the eigenvalues of $\tilde{\mathbf{r}}_{S}^{(S)}$ are respectively $A_0 A_{01}$, $-A_{z}$, $-A_0 A_{01}$; and those of $\tilde{\mathbf{t}}_{S}^{(S)}$ are respectively $(1 + A_0 A_{01})$, $(1 - A_{z})$, $k_{izS}^{(S)}(1 + A_0 A_{01})/k_{tzS}^{(S)}$.

The above-mentioned plane-of-incidence based local $S'$ coordinate systems are precisely the ones used for the TM-TE decomposition of the constituent plane-wave fields while using Fresnel coefficients (e.g. Ref. \cite{Berry435}). The above alternative derivation shows that the $\tilde{\mathbf{j'}}_S^{(S)}$ matrices conveniently include all information regarding these local coordinate systems. So, the direct application of the $\tilde{\mathbf{j'}}_S^{(S)}$ matrices automatically avoids any intermediate step involving these local coordinate systems --- thus contributing to the mathematical elegance of our formalism.

\subsection{Relation to Fresnel Coefficients}

In the $S'$ coordinate system, the field amplitude vectors $\boldsymbol{\mathcal{E}}_{jS}^{(S')}$ are easily decomposed into TM-TE components as
\begin{equation}
\boldsymbol{\mathcal{E}}_{jS}^{(S')} = \boldsymbol{\mathcal{E}}_{j(TM)S}^{(S')} + \boldsymbol{\mathcal{E}}_{j(TE)S}^{(S')};\label{EjSP=TE+TM}
\end{equation}
\vspace{-1.4em}
\begin{eqnarray}
\mbox{where,} \hspace{1em}
\boldsymbol{\mathcal{E}}_{j(TM)S}^{(S')} &=& \mathcal{E}_{jxS}^{(S')}\, \hat{\mathbf{x}}^{(S')} + \mathcal{E}_{jzS}^{(S')}\, \hat{\mathbf{z}}^{(S')};\label{EjSP_TM}\\
\boldsymbol{\mathcal{E}}_{j(TE)S}^{(S')} &=& \mathcal{E}_{jyS}^{(S')}\, \hat{\mathbf{y}}^{(S')}.\label{EjSP_TE}
\end{eqnarray}
Then, according to Eqs. (\ref{EjSSP}--\ref{tMatSP}), the 
TE components of the reflected and transmitted field amplitude vectors are given by
\begin{eqnarray}
\boldsymbol{\mathcal{E}}_{r(TE)S}^{(S')} &=& -A_z\, \boldsymbol{\mathcal{E}}_{i(TE)S}^{(S')}\,;\\
\boldsymbol{\mathcal{E}}_{t(TE)S}^{(S')} &=& (1 - A_z)\, \boldsymbol{\mathcal{E}}_{i(TE)S}^{(S')}\,.
\end{eqnarray}
It is easily verified by using Eq. (\ref{Axyz0}) that $r_{TE} = -A_z$ is the Fresnel TE reflection coefficient and $t_{TE} = 1 - A_z$ is the Fresnel TE transmission coefficient.

Now, if $\mathcal{E}_{j(TM)S}^{(S')}$ are the amplitude values of $\boldsymbol{\mathcal{E}}_{j(TM)S}^{(S')}$ with reference to the vector directions given in Refs. \cite{Jackson, BornWolf}, 
then the Fresnel TM reflection and transmission coefficients are respectively given by 
\begin{eqnarray}
r_{TM} &=& \mathcal{E}_{r(TM)S}^{(S')}/\mathcal{E}_{i(TM)S}^{(S')};\label{rTM}\\
t_{TM} &=& \mathcal{E}_{t(TM)S}^{(S')}/\mathcal{E}_{i(TM)S}^{(S')}.\label{tTM}
\end{eqnarray}
We have determined $\boldsymbol{\mathcal{E}}_{j'(TM)S}^{(S')}$ by using Eqs. (\ref{EjSSP}, \ref{EjSP_TM}); and have used them to verify that Eqs. (\ref{rTM}, \ref{tTM}) reduce to the standard Fresnel TM coefficient expressions given in Refs. \cite{Jackson, BornWolf}.
The same Eq. (\ref{rTM}) also gives the $r_{TM}$ expression of Ref. \cite{SalehTeich}, if $\mathcal{E}_{r(TM)S}^{(S')}$ is the amplitude value of $\boldsymbol{\mathcal{E}}_{r(TM)S}^{(S')}$ with reference to the corresponding vector direction given in the same Ref. \cite{SalehTeich}. However, since we always use Eq. (\ref{EjSS}--\ref{tMat}) in our formalism, without ever reducing them to Eqs. (\ref{EjSP=TE+TM}--\ref{tTM}), the sign-ambiguity of $r_{TM}$ is eliminated altogether.



\section{The Complete Formalism}\label{Sec_Formalism}


In an actual problem, a complete incident wave field $\tilde{\mathbf{E}}_{iI}^{(I)}$ is decomposed into constituent plane waves in terms of either Fourier decomposition or wavefront-surface-element decomposition. Fourier decomposition is a standard method, which we discuss qualitatively in Subsection \ref{Sub_Fourier}. For the purpose of the present work we use wavefront-surface-element decomposition in the simulation, whose mathematical description we give in Subsection \ref{Sub_WSE}. 


\subsection{The Incident Constituent Plane Waves : Fourier Decomposition}\label{Sub_Fourier}

\begin{figure}[t]
\centering
\includegraphics[width = 0.67\linewidth]{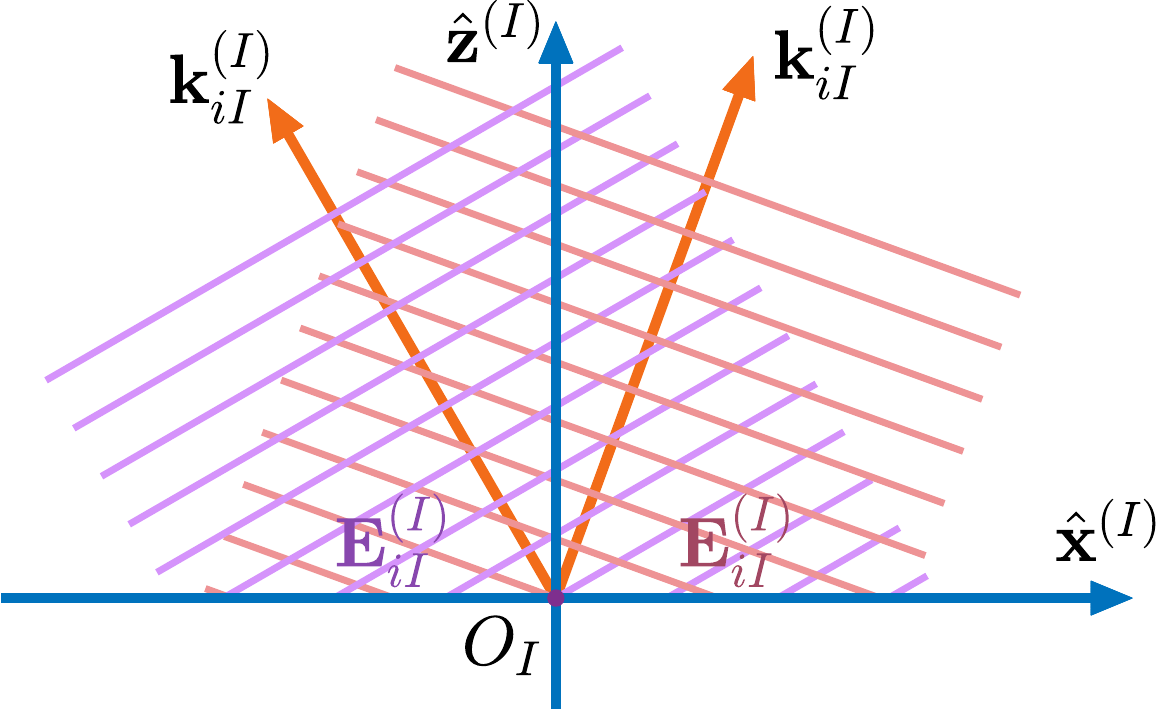}
\caption{\small{
A qualitative representation of the Fourier decomposition of a complete $\tilde{\mathbf{E}}_{iI}^{(I)}$ field. The field $\tilde{\mathbf{E}}_{iI}^{(I)}$ is decomposed into ideal $\mathbf{k}_{iI}^{(I)}$-$\mathbf{E}_{iI}^{(I)}$ plane waves. Each constituent plane-wave field $\mathbf{E}_{iI}^{(I)}$ exists everywhere at the surface $I$; and its phase-variation is obtained as a function of $\left( x^{(I)}, y^{(I)} \right)$.
}}
{\color{grey}{\rule{\linewidth}{1pt}}}
\label{fig_FourierComps}
\end{figure}

In the Fourier decomposition approach, the complete incident field $\tilde{\mathbf{E}}_{iI}^{(I)}$ is decomposed into Fourier component plane waves fields by using the formalism of spatial spectral analysis \cite{SalehTeich}. One can qualitatively visualize this decomposition in a way as shown in Fig. \ref{fig_FourierComps}. Infinitely many constituent plane waves, characterized by their wavevectors $\mathbf{k}_{iI}^{(I)}$, are obtained in the process. The electric field of each of these plane waves, at the surface $I$, has the form
\begin{equation}
\mathbf{E}_{iI}^{(I)} = \boldsymbol{\mathcal{E}}_{iI}^{(I)} e^{ i \left( \mathbf{k}_{iI}^{(I)} \cdot \Delta \mathbf{r}_{IO}^{(I)} \, -\, \omega t \, +\, \Phi_{O} \right) }.\label{EiII_Fourier}
\end{equation}
Here, $\boldsymbol{\mathcal{E}}_{iI}^{(I)}$ is the field amplitude vector that contains the polarization information. It is a constant not only at the surface $I$, but everywhere in the concerned constituent wave. 

The phase term $\Phi_O$ is the phase of the constituent wave at a pre-assigned global reference point. For example, we can assign the origin $O_I$ to be the global reference point, with respect to which all path-dependent phases in the entire system can be determined. 

The term $\Delta\mathbf{r}_{IO}^{(I)}$ is the position vector of a concerned point $\left(x^{(I)},y^{(I)},0^{(I)}\right)$ with respect to the above-mentioned pre-assigned reference point. For example, if the origin $O_I$ is the reference point, then $\Delta\mathbf{r}_{IO}^{(I)}$ is simply the position vector $\Delta\mathbf{r}_{IO}^{(I)} = x^{(I)} \, \hat{\mathbf{x}}^{(I)} + y^{(I)} \, \hat{\mathbf{y}}^{(I)}$. So, the term $\mathbf{k}_{iI}^{(I)} \cdot \Delta\mathbf{r}_{IO}^{(I)}$ is the path-dependent phase of the concerned constituent plane wave at $\left(x^{(I)},y^{(I)},0^{(I)}\right)$, with respect to the pre-assigned reference phase at $O_I$. It is to be noticed that $\Delta\mathbf{r}_{IO}^{(I)}$ is the only term in Eq. (\ref{EiII_Fourier}) which is a function of $\left(x^{(I)}, y^{(I)}\right)$. The spatial variation of the constituent plane-wave field $\mathbf{E}_{iI}^{(I)}$ as a function of $\left(x^{(I)}, y^{(I)}\right)$ is thus completely determined at any given time $t$ by using Eq. (\ref{EiII_Fourier}).


\subsection{The Incident Constituent Plane Waves : Wavefront-Surface-Element Decomposition}\label{Sub_WSE}

For incident waves with spherical, cylindrical and other simple geometrically-shaped wavefronts, the method of wavefront-surface-element decomposition is convenient. We describe this method by directly taking the example of our simulated wave model, which involves a spherical incident wave.

In this subsection, we first demonstrate the construction of our wave model. Subsequently we explain how the wavefront-surface-element decomposition is carried out in this model.

\vspace{1em}

\begin{figure}[t]
\centering
\includegraphics[width = 0.72487\linewidth]{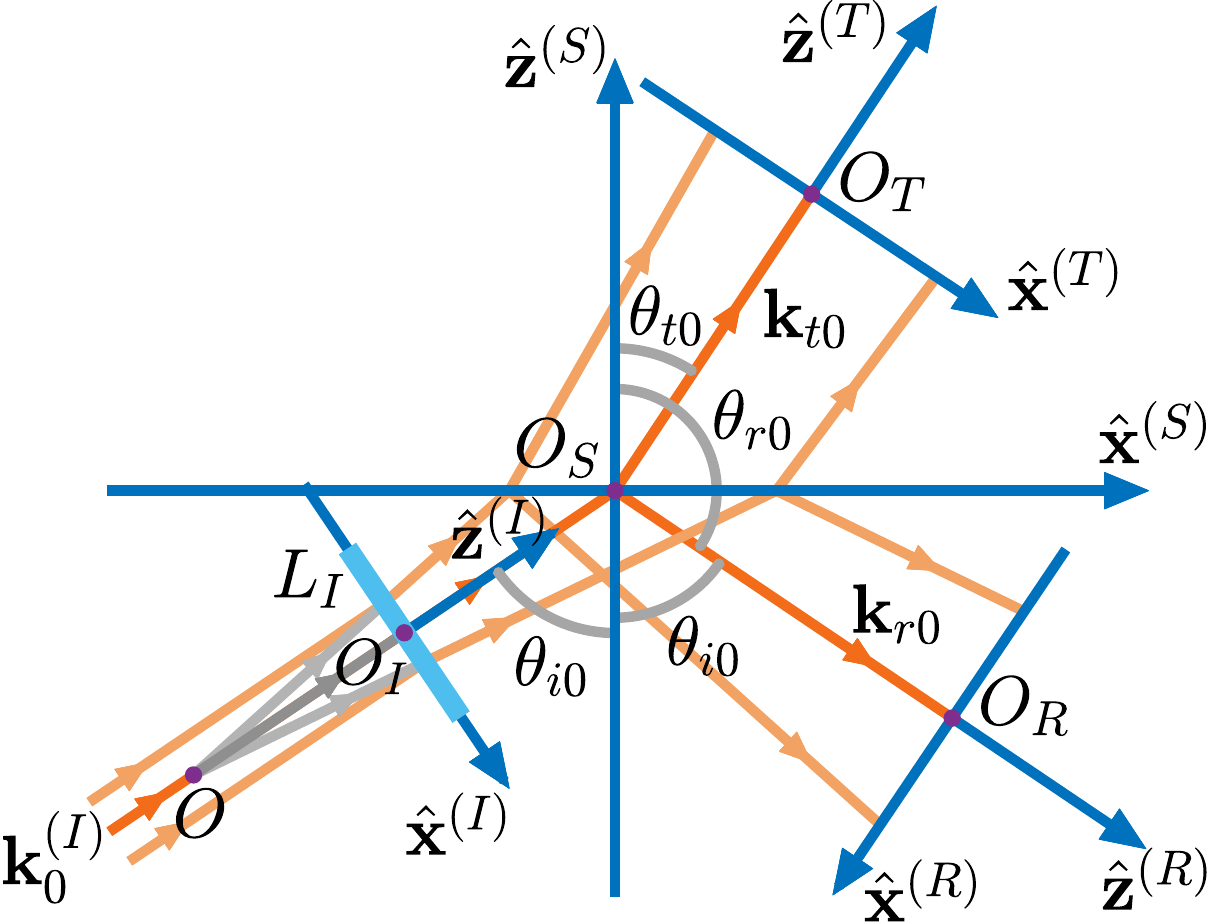}
\caption{\small{The simulated diverging beam model. A collimated beam is diverged through the lens $L_I$. The complete incident electric field $\tilde{\mathbf{E}}_{iI}^{(I)}$ is determined analytically; and the complete reflected and transmitted electric fields $\tilde{\mathbf{E}}_{j'J'}^{(J')}$ are determined computationally.}}
{\color{grey}{\rule{\linewidth}{1pt}}}
\label{fig_setup}
\end{figure}

\textbf{Model Construction:} The model is shown in Fig. \ref{fig_setup}. An initial plane-wave beam is considered having a wavevector
\begin{equation}
\mathbf{k}_{0}^{(I)} = n_1 k\, \hat{\mathbf{z}}^{(I)} \label{k0I=n1kz}
\end{equation}
($k = 2\pi/\lambda$; $\lambda = $ free space wavelength) and an electric field $\mathbf{E}_0^{(I)}$ with the amplitude vector profile (Gaussian)
\begin{eqnarray}
&\boldsymbol{\mathcal{E}}_0^{(I)} 
= \boldsymbol{\mathcal{E}}_{0x}^{(I)} + e^{i\Phi_E} \boldsymbol{\mathcal{E}}_{0y}^{(I)} = \mathcal{E}_{0x}^{(I)} \hat{\mathbf{x}}^{(I)} + e^{i\Phi_E} \mathcal{E}_{0y}^{(I)} \hat{\mathbf{y}}^{(I)}; \hspace{1em} & \label{E0I} \\
&\mathcal{E}_{0x}^{(I)}  \equiv \mathcal{E}_{0x}^{(I)} \left( x^{(I)}, y^{(I)} \right) = \mathcal{E}_{00}\, e^{-\rho^{(I)\, 2}/w_0^2} \cos\theta_E; \hspace{1em} & \label{E0xI} \\
&\mathcal{E}_{0y}^{(I)} \equiv \mathcal{E}_{0y}^{(I)} \left( x^{(I)}, y^{(I)} \right) = \mathcal{E}_{00}\, e^{-\rho^{(I)\, 2}/w_0^2} \sin\theta_E; \hspace{1em} & \label{E0yI} \\
& \mbox{where,} \hspace{1em} \rho^{(I)} = \left( x^{(I)\,2} + y^{(I)\,2} \right)^{\frac{1}{2}}; & \label{rhoI}
\end{eqnarray}
where, $\mathcal{E}_{00} $ is the central electric field magnitude; $w_0 $ is the half beam-width; $\theta_E,\Phi_E $ are fixed angle and relative phase terms which determine the polarization of $\boldsymbol{\mathcal{E}}_{0}^{(I)}$. The plane-wave beam passes through a concave lens $L_I$ of focal length $f$ ($<0$), placed at the surface $I$, centered at $O_I$; by which it is converted to a spherically diverging beam with an angle of divergence $2\theta_D = -2\tan^{-1}(w_0/f)$ and with the center of curvature at point $O$ (focus of $L_I$; $f = -OO_I$), as shown in Fig. \ref{fig_setup}. This diverging beam, with complete field function $\tilde{\mathbf{E}}_{iI}^{(I)}$ (calculated later in this subsection), serves the purpose of the incident complex optical wave in our model.

\vspace{1em}

\begin{figure}[t]
\centering
\includegraphics[width = 0.796\linewidth]{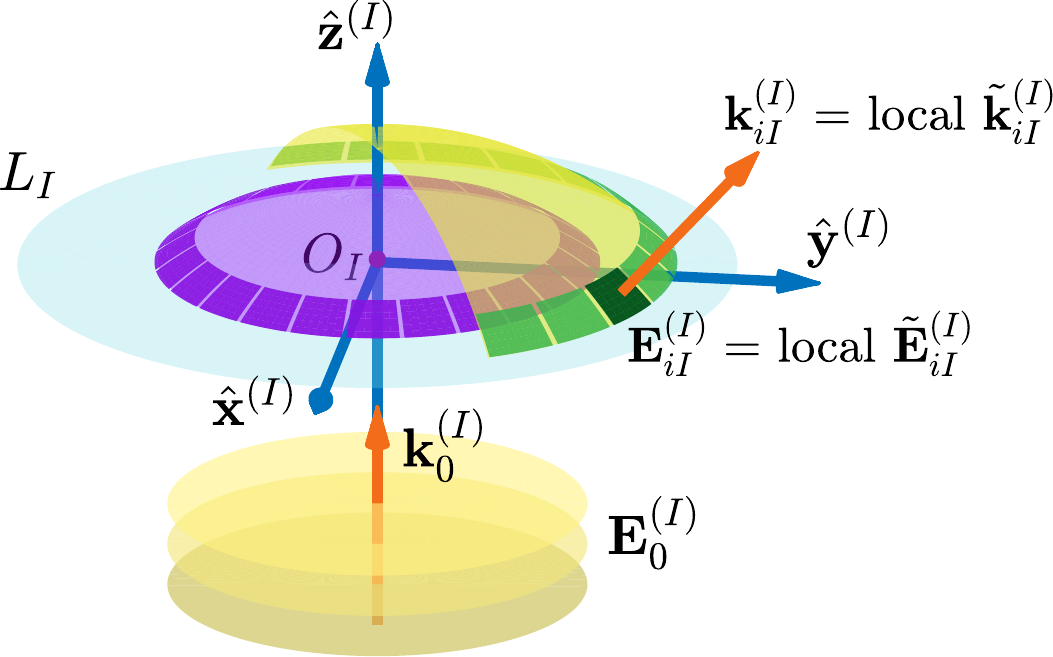}
\caption{\small{
The $\mathbf{k}_0^{(I)}$-$\mathbf{E}_0^{(I)}$ plane-wave beam is converted to the spherically diverging $\tilde{\mathbf{k}}_{iI}^{(I)}$-$\tilde{\mathbf{E}}_{iI}^{(I)}$ beam by the lens $L_I$. The surface $I$ (the plane of $L_I$) intercepts each spherical wavefront of the $\tilde{\mathbf{E}}_{iI}^{(I)}$ field at a circular boundary, where, surface-elements of the $\tilde{\mathbf{E}}_{iI}^{(I)}$ wavefront are considered. Each such surface element acts as a constituent $\mathbf{k}_{iI}^{(I)}$-$\mathbf{E}_{iI}^{(I)}$ plane wave.
}}
{\color{grey}{\rule{\linewidth}{1pt}}}
\label{fig_WSEComps}
\end{figure}

\textbf{Spherical Wavefronts and their Decomposition:}
One can visualize the transformation of the field $\mathbf{E}_0^{(I)}$ to the field $\tilde{\mathbf{E}}_{iI}^{(I)}$ in a way as shown in Fig. \ref{fig_WSEComps}. The plane wavefronts of $\mathbf{E}_0^{(I)}$ are transformed to spherical wavefronts of $\tilde{\mathbf{E}}_{iI}^{(I)}$ by the lens $L_I$ at the surface $I$. For convenience, we consider $L_I$ to have an infinitesimal thickness. Figure \ref{fig_WSEComps} shows, as examples, two such spherical wavefronts (one fully and another partially). Each such wavefront is intercepted at the surface $I$ at a circular boundary. We consider the surface elements of each $\tilde{\mathbf{E}}_{iI}^{(I)}$ wavefront at this circular intercept region only --- as shown in Fig. \ref{fig_WSEComps} --- so that 
they are identified by coordinates $\left(x^{(I)}, y^{(I)},0^{(I)}\right)$. Considering such sets of surface elements for many wavefronts of $\tilde{\mathbf{E}}_{iI}^{(I)}$ everywhere on the surface $I$, the wavefront-surface-element decomposition of $\tilde{\mathbf{E}}_{iI}^{(I)}$ is achieved.

Based on the above qualitative picture, we now mathematically analyse the decomposition process. We consider a spherical wavefront, whose circular intercept at the surface $I$ has a radius $\rho^{(I)}$. The radius of curvature of this wavefront 
is then given by
\begin{eqnarray}
& r_I 
= \left(\rho^{(I)\, 2} + f^2\right)^{\frac{1}{2}}.& \label{rI_def}
\end{eqnarray}
If a concerned surface element of this wavefront is located at $\left(x^{(I)}, y^{(I)},0^{(I)}\right)$, then $\left(x^{(I) \; 2} + y^{(I) \; 2}\right)^{1/2} = \rho^{(I)}$ [Eq. (\ref{rhoI})]. 
We define
\begin{eqnarray}
x^{(I)}/\rho^{(I)} = \cos\phi_I; &\hspace{1em}&
y^{(I)}/\rho^{(I)} = \sin\phi_I; \label{cs_phiI} \\
|f|/r_I = \cos\theta_I; &\hspace{1em}&
\rho^{(I)}/r_I = \sin\theta_I. \label{cs_thetaI}
\end{eqnarray}
Then, the wavevector 
at the concerned surface element is given by
\begin{eqnarray}
& \tilde{\mathbf{k}}_{iI}^{(I)} \left(x^{(I)}, y^{(I)}\right) = n_1 k \, \hat{\mathbf{r}}_{I}^{(I)}; & \label{kiII=n1kr} \\
& \hat{\mathbf{r}}_{I}^{(I)} \! = \sin\theta_I \cos\phi_I \, \hat{\mathbf{x}}^{(I)} \! + \sin\theta_I \sin\phi_I \, \hat{\mathbf{y}}^{(I)} \! + \cos\theta_I \, \hat{\mathbf{z}}^{(I)}; \hspace{1.5em} & \label{rII}
\end{eqnarray}
where, $\hat{\mathbf{r}}_{I}^{(I)}$ is the unit vector normal to the 
surface element. The complete electric field amplitude vector 
at the surface element is given by [Appendix \ref{App_Ei_Model}]
\begin{eqnarray}
&\tilde{\boldsymbol{ \mathcal{E}}}_{iI}^{(I)} \left(x^{(I)}, y^{(I)}\right) = g_I\,  \tilde{\mathbf{R}}_{II'} \tilde{\mathbf{R}}_{I'I''} \tilde{\mathbf{R}}_{I'I} \, \boldsymbol{\mathcal{E}}_{0}^{(I)};\hspace{1.5em} & \label{EiII_model} \\
&\mbox{where,} \hspace{0.5em} \tilde{\mathbf{R}}_{I'I} = 
\begin{bmatrix}
\cos\phi_I & \sin\phi_I & 0 \\
-\sin\phi_I & \cos\phi_I & 0 \\
0 & 0 & 1
\end{bmatrix} \! ;
\hspace{0.4em} \tilde{\mathbf{R}}_{II'} = \tilde{\mathbf{R}}_{I'I}^{-1} \, ;\hspace{1.5em}& \label{RIIP} \\
&\tilde{\mathbf{R}}_{I'I''} = 
\begin{bmatrix}
\cos\theta_I & 0 & \sin\theta_I \\
0 & 1 & 0 \\
-\sin\theta_I & 0 & \cos\theta_I
\end{bmatrix} \! ;
\hspace{0.5em} g_I = \dfrac{1}{\sqrt{\cos\theta_I}} \, . \hspace{1em} 
& \label{RIPPgI}
\end{eqnarray}

The path-dependent phase of the field $\tilde{\mathbf{E}}_{iI}^{(I)}$ at the concerned surface element is determined based on the geometry of the spherical wavefront. We assign a reference phase $\Phi_O = 0$ to the center of curvature point $O$ [Fig. \ref{fig_setup}] \cite{FOOTNOTE_VirtualSource}. Then, using Eqs. (\ref{rI_def}, \ref{kiII=n1kr}, \ref{rII}), the phase 
of the field $\tilde{\mathbf{E}}_{iI}^{(I)}$ at the surface element is obtained as 
\begin{eqnarray}
& \tilde{\Phi}_{iI} \left(x^{(I)}, y^{(I)}\right) = \tilde{\mathbf{k}}_{iI}^{(I)} \cdot r_I \, \hat{\mathbf{r}}_I^{(I)} = n_1 k \, r_I \, . & \label{Phi_iI}
\end{eqnarray}

In the above discussion we have considered a general surface element located at $\left(x^{(I)}, y^{(I)}, 0^{(I)}\right)$. So, Eqs. (\ref{kiII=n1kr}, \ref{EiII_model}, \ref{Phi_iI}) give the functional forms of the wavevector, field amplitude vector and phase of the complete incident field in terms of the coordinates $\left(x^{(I)}, y^{(I)}\right)$. Now, each such surface element is considered as a plane wave in the present decomposition method. Then, the local values of the quantities 
$\tilde{\mathbf{k}}_{iI}^{(I)}$, $\tilde{\boldsymbol{ \mathcal{E}}}_{iI}^{(I)}$ and $\tilde{\Phi}_{iI}$ at $\left(x^{(I)}, y^{(I)}\right)$ serve the purposes of the plane wavevector $\mathbf{k}_{iI}^{(I)}$, the corresponding electric field amplitude vector $\boldsymbol{\mathcal{E}}_{iI}^{(I)}$ and the path-dependent phase $\mathbf{k}_{iI}^{(I)}\cdot \Delta\mathbf{r}_{IO}^{(I)}$ 
of the considered surface-element plane-wave; i.e. 
\begin{subequations}
\label{kiII,EiII,PhiI=local}
\begin{eqnarray}
\mathbf{k}_{iI}^{(I)} &=& \,\mbox{local } \tilde{\mathbf{k}}_{iI}^{(I)} \left(x^{(I)}, y^{(I)}\right); \label{kiII=local} \\ 
\boldsymbol{\mathcal{E}}_{iI}^{(I)} &=& \,\mbox{local } \tilde{\boldsymbol{ \mathcal{E}}}_{iI}^{(I)} \left(x^{(I)}, y^{(I)}\right); \label{EiII=local} \\
\mathbf{k}_{iI}^{(I)} \cdot \Delta\mathbf{r}_{IO}^{(I)} &=& \,\mbox{local } \tilde{\Phi}_{iI} \left(x^{(I)}, y^{(I)}\right). \label{PhiI=local}
\end{eqnarray}
\end{subequations}
In particular, at $\left(x^{(I)}, y^{(I)}\right) = (0,0)$, we get the central wavevector
\begin{equation}
\mathbf{k}_{i0} = \tilde{\mathbf{k}}_{iI}^{(I)} (0,0) = \mathbf{k}_{0}^{(I)}, \hspace{1em} [\mbox{Eq. (\ref{k0I=n1kz})}].
\end{equation}

Using the terms in Eqs. (\ref{kiII,EiII,PhiI=local}) and $\Phi_O = 0$, the electric field of the concerned surface-element plane-wave is obtained as
\begin{equation}
\mathbf{E}_{iI}^{(I)} = \boldsymbol{\mathcal{E}}_{iI}^{(I)} e^{ i \left( \mathbf{k}_{iI}^{(I)} \cdot \Delta \mathbf{r}_{IO}^{(I)} \, -\, \omega t \, +\, \Phi_{O} \right) };\label{EiII_WSE}
\end{equation}
which is the exact same equation as Eq. (\ref{EiII_Fourier}). The mathematical formulation of the intended wavefront-surface-element decomposition is thus achieved. The complete field $\tilde{\mathbf{E}}_{iI}^{(I)}$ is simply the collection of all these local surface-element fields; and is expressed in the functional form
\begin{equation}
\tilde{\mathbf{E}}_{iI}^{(I)} = \tilde{\boldsymbol{\mathcal{E}}}_{iI}^{(I)} e^{ i \left( \tilde{\Phi}_{iI} \, -\, \omega t \, +\, \Phi_{O} \right) }.\label{EiII_WSE_full}
\end{equation}

\vspace{2em}

The surface-element decomposition of cylindrical and other simple-shaped wavefronts are also achieved in a similar way by using appropriate geometry of the concerned system.
For convenience, we describe the rest of the formalism in Subsections \ref{Sub_Propagation} -- \ref{Sub_Recombination} by assuming surface-element decomposition; and then summarize the equivalent description of the formalism in terms of Fourier decomposition in Subsection \ref{Sub_FourierRecombination}.





\subsection{Propagation from Surface I to Surface S}\label{Sub_Propagation}

Starting from the position $\left( x^{(I)}, y^{(I)}, 0^{(I)} \right)$, the concerned surface-element wave propagates along $\mathbf{k}_{iI}^{(I)}$ and reaches a point $\left( x^{(S)}, y^{(S)}, 0^{(S)} \right)$ at the surface $S$ [Fig. \ref{fig_setup}]. Since the direction of this propagation is constant, we get $\mathbf{k}_{iS}^{(I)} = \mathbf{k}_{iI}^{(I)}$. Also, we denote this displacement of the surface element by a vector $\Delta \mathbf{r}_{SI}^{(I)}$, which is easily calculated from the geometry of the system. Then, the change in phase of the surface-element plane-wave field due to this propagation is given by $\Phi_{SI} = \mathbf{k}_{iI}^{(I)} \cdot \Delta \mathbf{r}_{SI}^{(I)}$.

It is to be noticed that, the planar approximation of the surface element is valid on the wavefront because the physical dimensions of the element are very small as compared to the radius of curvature of the wavefront. However, when the surface element propagates to large distances comparable to the wavefront's radius of curvature, the resulting change in size of the element must be taken into account.
For example, in our spherical wavefront model, this change in size directly results into the inverse-square variation of the field intensity. This implies that, $\boldsymbol{\mathcal{E}}_{iI}^{(I)}$ changes by a factor $g_{SI} = r_I/r_S$, where $r_I$ and $r_S$ are the radii of curvature of the surface element at the coordinates $\left( x^{(I)}, y^{(I)}, 0^{(I)} \right)$ and $\left( x^{(S)}, y^{(S)}, 0^{(S)} \right)$ respectively. Similar $g_{SI}$ expressions can be calculated for other simple-shaped wavefronts as well. The field amplitude vector at $\left( x^{(S)}, y^{(S)}, 0^{(S)} \right)$ thus becomes $\boldsymbol{\mathcal{E}}_{iS}^{(I)} = g_{SI}\, \boldsymbol{\mathcal{E}}_{iI}^{(I)}$. In this way, the factor $g_{SI}$ inherently contains the complete information about the wavefront curvature, even when the surface element is locally considered as a plane wave.

Considering the above expressions, the incident surface-element plane-wave field at $\left( x^{(S)}, y^{(S)}, 0^{(S)} \right)$ is obtained as $\mathbf{E}_{iS}^{(I)} = g_{SI} \mathbf{E}_{iI}^{(I)} e^{i\Phi_{SI}}$.

Now, $\mathbf{k}_{iS}^{(I)}$ and $\mathbf{E}_{iS}^{(I)}$ are expressed in terms of the $S$ coordinate system by applying the rotation matrix
\begin{equation}
\tilde{\mathbf{R}}_{SI} = 
\begin{bmatrix}
\cos\theta_{i0} & 0 & \sin\theta_{i0}\\
0 & 1 & 0\\
-\sin\theta_{i0} & 0 & \cos\theta_{i0}
\end{bmatrix}.
\end{equation}
We thus obtain
\begin{eqnarray}
&\mathbf{k}_{iS}^{(S)} = \tilde{\mathbf{R}}_{SI} \mathbf{k}_{iS}^{(I)} = \tilde{\mathbf{R}}_{SI} \mathbf{k}_{iI}^{(I)};\label{kiSS}
&\\
&\mathbf{E}_{iS}^{(S)} = \tilde{\mathbf{R}}_{SI} \mathbf{E}_{iS}^{(I)} = g_{SI} \tilde{\mathbf{R}}_{SI} \mathbf{E}_{iI}^{(I)} e^{i\Phi_{SI}}.&\label{EiSS}
\end{eqnarray}

\subsection{Reflection and Transmission}

Equation (\ref{kiSS}) gives the incident wavevector $\mathbf{k}_{iS}^{(S)}$, which is used in Eq. (\ref{kjSS}) to obtain the reflected and transmitted wavevectors 
\begin{equation}
\mathbf{k}_{j'S}^{(S)} = \tilde{\mathbf{I}}_{j'S}^{(S)}\, \mathbf{k}_{iS}^{(S)} = \tilde{\mathbf{I}}_{j'S}^{(S)}\, \tilde{\mathbf{R}}_{SI} \mathbf{k}_{iI}^{(I)}.\label{kjSS_B}
\end{equation}
Equation (\ref{EiSS}) gives the incident electric field $\mathbf{E}_{iS}^{(S)}$, on which the $\tilde{\mathbf{j'}}_{S}^{(S)}$ matrices [Eqs. (\ref{rMat}, \ref{tMat})] are applied to obtain the reflected and transmitted electric fields 
\begin{equation}
\mathbf{E}_{j'S}^{(S)} = \tilde{\mathbf{j'}}_{S}^{(S)} \mathbf{E}_{iS}^{(S)} = g_{SI}\, \tilde{\mathbf{j'}}_{S}^{(S)} \tilde{\mathbf{R}}_{SI} \mathbf{E}_{iI}^{(I)} e^{i \Phi_{SI}}.
\label{EjSS_B}
\end{equation}
We thus obtain two new surface-element plane-waves --- one reflected and the other transmitted --- corresponding to the incident surface-element wave at the same position $\left( x^{(S)}, y^{(S)}, 0^{(S)} \right)$. Their geometrical properties are comparable to those of the $\mathbf{E}_{iI}^{(I)}$ surface-element [Eq. (\ref{EiII_WSE})] considered at $\left( x^{(I)}, y^{(I)}, 0^{(I)} \right)$.

\subsection{The Reflected Wavevector and Field}

The reflected field is observed at the surface $R$. The propagation of the reflected surface-element wave from the surface $S$ to the surface $R$ is geometrically similar to the propagation of the incident surface-element wave from the surface $I$ to the surface $S$. From the position $\left( x^{(S)}, y^{(S)}, 0^{(S)} \right)$, the reflected surface-element wave propagates along $\mathbf{k}_{rS}^{(S)}$ to reach the point $\left( x^{(R)}, y^{(R)}, 0^{(R)} \right)$, which is determined geometrically, with the displacement vector being denoted by $\Delta\mathbf{r}_{RS}^{(S)}$. Since the propagation direction is unchanged, we get $\mathbf{k}_{rR}^{(S)} = \mathbf{k}_{rS}^{(S)}$.

The reflected field amplitude gets modified by a multiplicative factor $g_{RS}$, which is geometrically similar in nature to the factor $g_{SI}$ of Eq. (\ref{EiSS}).
The field is also modified by a path-dependent phase $\Phi_{RS} = \mathbf{k}_{rS}^{(S)} \cdot \Delta\mathbf{r}_{RS}^{(S)}$ due to the propagation. Thus, the reflected surface-element plane-wave field at $\left( x^{(R)}, y^{(R)}, 0^{(R)} \right)$ is given by $\mathbf{E}_{rR}^{(S)} = g_{RS} \mathbf{E}_{rS}^{(S)} e^{i\Phi_{RS}}$.

Finally, $\mathbf{k}_{rR}^{(S)}$ and $\mathbf{E}_{rR}^{(S)}$ are expressed in terms of the $R$ coordinate system by applying the rotation matrix
\begin{equation}
\tilde{\mathbf{R}}_{RS} = 
\begin{bmatrix}
\cos\theta_{r0} & 0 & -\sin\theta_{r0}\\
0 & 1 & 0\\
\sin\theta_{r0} & 0 & \cos\theta_{r0}
\end{bmatrix}.
\end{equation}
We thus obtain
\begin{eqnarray}
&& \hspace{0.9em} \mathbf{k}_{rR}^{(R)} = \tilde{\mathbf{R}}_{RS} \mathbf{k}_{rR}^{(S)} = \tilde{\mathbf{R}}_{RS} \tilde{\mathbf{I}}_{rS}^{(S)}\, \tilde{\mathbf{R}}_{SI} \mathbf{k}_{iI}^{(I)}; \label{krRR} \\
&& \hspace{-1em} \mathbf{E}_{rR}^{(R)} = \tilde{\mathbf{R}}_{RS} \mathbf{E}_{rR}^{(S)}  \nonumber\\
&& \hspace{1.315em} = g_{RS}\, g_{SI} \tilde{\mathbf{R}}_{RS}\, \tilde{\mathbf{r}}_{S}^{(S)} \tilde{\mathbf{R}}_{SI} \mathbf{E}_{iI}^{(I)} e^{i \left(\Phi_{RS} + \Phi_{SI}\right)}.\label{ErRR}
\end{eqnarray}

\subsection{The Transmitted Wavevector and Field}

The transmitted surface-element plane wave propagates from $\left( x^{(S)}, y^{(S)}, 0^{(S)} \right)$ to a corresponding position $\left( x^{(T)}, y^{(T)}, 0^{(T)} \right)$ in the same way as the reflected surface-element plane wave propagates from $\left( x^{(S)}, y^{(S)}, 0^{(S)} \right)$ to $\left( x^{(R)}, y^{(R)}, 0^{(R)} \right)$ (a note on the relevant amplitude modifying multiplicative factor $g_{TS}$ is given in Appendix \ref{App_gTS}). So, we can write the transmitted wavevector and surface-element electric field expressions at the position $\left( x^{(T)}, y^{(T)}, 0^{(T)} \right)$ simply by replacing the scripts $r\rightarrow t$ and $R\rightarrow T$ in Eqs. (\ref{krRR}, \ref{ErRR}). Thus, we get
\begin{eqnarray}
&\mathbf{k}_{tT}^{(T)} = \tilde{\mathbf{R}}_{TS} \mathbf{k}_{tT}^{(S)} 
= \tilde{\mathbf{R}}_{TS} \tilde{\mathbf{I}}_{tS}^{(S)}\, \tilde{\mathbf{R}}_{SI} \mathbf{k}_{iI}^{(I)};&\label{ktTT}
\\
&\mathbf{E}_{tT}^{(T)}  
= g_{TS}\, g_{SI} \tilde{\mathbf{R}}_{TS}\, \tilde{\mathbf{t}}_{S}^{(S)} \tilde{\mathbf{R}}_{SI} \mathbf{E}_{iI}^{(I)} e^{i \left(\Phi_{TS} + \Phi_{SI}\right)}.& \label{EtTT}
\end{eqnarray}


\subsection{Recombination of the Plane Waves}\label{Sub_Recombination}

Equations (\ref{ErRR}, \ref{EtTT}) 
give the final surface-element plane-wave field $\mathbf{E}_{j' J'}^{(J')}$ at the position $\left( x^{(J')}, y^{(J')}, 0^{(J')} \right)$ at each observing surface $J'$ ($ = R,T$).  
The collection of all the surface-element fields over the entire surface $J'$ 
gives the complete reflected/transmitted electric field $\tilde{\mathbf{E}}_{j'J'}^{(J')}$ as a function of $\left( x^{(J')}, y^{(J')} \right)$. 
In fact, the same process can also be applied to combine the surface-element plane-wave fields $\mathbf{E}_{jS}^{(S)}$ [Eqs. (\ref{EiSS}, \ref{EjSS_B})] to obtain the complete field $\tilde{\mathbf{E}}_{jS}^{(S)}$ at the surface $S$, as a function of $\left( x^{(S)}, y^{(S)} \right)$.

\subsection{The Use of Fourier Decomposition}\label{Sub_FourierRecombination}

The equivalence of Eqs. (\ref{EiII_Fourier}, \ref{EiII_WSE}) is readily extended to the rest of the formalism as well. Equations (\ref{kiSS}--\ref{EtTT}) are readily applicable to the Fourier decomposition case with proper reinterpretation of the various terms. Instead of being a point-to-point displacement, 
each term $\Delta\mathbf{r}_{C_2 C_1}^{(C_1)}$ now represents a displacement function to everywhere at the surface $C_2$ from a chosen reference point at the surface $C_1$ (e.g. the origin $O_{C_1}$). Consequently, each constituent plane-wave field $\mathbf{E}_{jC}^{(C)}$ [Eqs. (\ref{EiSS}, \ref{EjSS_B}, \ref{ErRR}, \ref{EtTT})] now becomes a plane-wave field existing over the entire surface $C$, as a function of $\left( x^{(C)}, y^{(C)} \right)$ (comparable to $\mathbf{E}_{iI}^{(I)} \left( x^{(I)}, y^{(I)} \right)$ of Eq. (\ref{EiII_Fourier}) and Fig. \ref{fig_FourierComps}).

Since the Fourier-decomposed plane waves are ideal plane waves, there is no change in area of the wavefronts due to propagation. So, the amplitude modifying factors $g_{SI}$, $g_{RS}$ and $g_{TS}$ are unity. Finally, the generic $\tilde{\mathbf{j'}}_S^{(S)}$ matrices [Eq. (\ref{rMat}, \ref{tMat})] are applicable to any incident plane-wave field irrespective of the considered decomposition method. Hence, with the above reinterpretations, Eqs. (\ref{EiSS}--\ref{EtTT}) are readily applicable to the case of Fourier decomposition to determine the reflected and transmitted constituent plane-wave fields.

However, the recombination methods of the constituent $\mathbf{E}_{j' J'}^{(J')}$ fields are different for the two decomposition cases. Since all $\mathbf{E}_{j' J'}^{(J')}$ fields exist everywhere at the surface $J'$ in the Fourier decomposition case, their superposition is performed all over the surface $J'$ to recombine them (a note on the geometric phase consideration is given in Appendix \ref{App_GeoPhase}). In this way, the complete electric field $\tilde{\mathbf{E}}_{j' J'}^{(J')}$ is obtained in the Fourier decomposition case.



\section{Simulation and Results}\label{Sec_Simulation}

In this section, we use short phrases such as `field $\boldsymbol{\mathcal{E}}$' and `field $\tilde{\boldsymbol{\mathcal{E}}}$' as per convenience --- instead of extended phrases such as `field amplitude vector $\boldsymbol{\mathcal{E}}$' and `field amplitude vector function $\tilde{\boldsymbol{\mathcal{E}}}$' --- for the simplicity of the discussions. The intended meanings are clearly understood by the notations $\boldsymbol{\mathcal{E}}$ and $\tilde{\boldsymbol{\mathcal{E}}}$.

We simulate the optical system of Fig. \ref{fig_setup} --- involving wavefront-surface-element decomposition --- and follow the steps described in Section \ref{Sec_Formalism} to generate the reflected and transmitted fields. 
Though our formalism is directly applicable to complex field amplitude vectors, it is computationally convenient to transform the linearly polarized $\boldsymbol{\mathcal{E}}_{0x}^{(I)}$ and $\boldsymbol{\mathcal{E}}_{0y}^{(I)}$ fields of Eq. (\ref{E0I}) 
separately in the simulation; and then superpose the individual output fields along with considering the phase difference $\Phi_E$ to obtain the complete transformations of $\boldsymbol{\mathcal{E}}_{0}^{(I)}$ [Eq. (\ref{E0I})]. 


With respect to the central plane of incidence (the $\mathbf{k}_{i0}$-$\hat{\mathbf{z}}^{(S)}$ plane), $\boldsymbol{\mathcal{E}}_{0x}^{(I)}$ is a TM polarized field and $\boldsymbol{\mathcal{E}}_{0y}^{(I)}$ is a TE polarized field. The transformed fields are neither TM nor TE, since they are distorted due to the curvature of the wavefronts [Eq. (\ref{EiII_model})]. However, in this section we use superscripts $X,Y$ with relevant quantities to indicate that the considered quantities correspond either to the initial TM field $\boldsymbol{\mathcal{E}}_{0x}^{(I)}$ (superscript $X$) or to the initial TE field $\boldsymbol{\mathcal{E}}_{0y}^{(I)}$ (superscript $Y$). 
Quantities without these superscripts are quantities corresponding to the total initial input field $\boldsymbol{\mathcal{E}}_{0}^{(I)}$.

\subsection{Data Grids}\label{Sub_DataGrids}

For computational results, we first generate grids of coordinate points $\left( x^{(C)},y^{(C)} \right)$ at all surfaces $C = S,J$. Based on each of these grids, we then computationally generate a complete set of relevant functions in the form of data grids. To explain what data is to be generated, we consider the example of the field $\tilde{\mathbf{E}}_{iI}^{(I)}$ [Eq. (\ref{EiII_WSE_full})]. By applying the transformation of Eq. (\ref{EiII_model}) to the fields $\boldsymbol{\mathcal{E}}_{0x}^{(I)}$ and $\boldsymbol{\mathcal{E}}_{0y}^{(I)}$, we obtain transformed fields which we denote respectively by $\tilde{\boldsymbol{\mathcal{E}}}_{iI}^{(I)X}$ and $\tilde{\boldsymbol{\mathcal{E}}}_{iI}^{(I)Y}$. Both $\tilde{\boldsymbol{\mathcal{E}}}_{iI}^{(I)X}$ and $\tilde{\boldsymbol{\mathcal{E}}}_{iI}^{(I)Y}$ are locally linearly polarized; but the linear polarization direction varies all over the surface $I$ due to the wavefront-curvature.

Now, according to Eq. (\ref{E0I}), the field $\boldsymbol{\mathcal{E}}_{0y}^{(I)}$ is first given an additional phase $\Phi_E$, and then superposed with the field $\boldsymbol{\mathcal{E}}_{0x}^{(I)}$ to obtain the complete initial input field $\boldsymbol{\mathcal{E}}_{0}^{(I)}$. Subsequently, by giving the field $\tilde{\boldsymbol{\mathcal{E}}}_{iI}^{(I)Y}$ the same additional phase $\Phi_E$, and then superposing it with the field $\tilde{\boldsymbol{\mathcal{E}}}_{iI}^{(I)X}$, we obtain the complete field $\tilde{\boldsymbol{\mathcal{E}}}_{iI}^{(I)}$ of Eq. (\ref{EiII_WSE_full}). Also, the phase function $\tilde{\Phi}_{iI}$ of Eq. (\ref{EiII_WSE_full}) is obtained in the form of Eq. (\ref{Phi_iI}) by using the wavevector-function $\tilde{\mathbf{k}}_{iI}^{(I)}$ of Eq. (\ref{kiII=n1kr}). 
Hence, the list of functional data grids to be computed at the surface $I$, to generate the complete field profile $\tilde{\mathbf{E}}_{iI}^{(I)}$ of Eq. (\ref{EiII_WSE_full}), is:




\vspace{0.5em}

$\tilde{k}_{ixI}^{(I)}$, 
$\tilde{k}_{iyI}^{(I)}$, 
$\tilde{k}_{izI}^{(I)}$, 
$\tilde{\mathcal{E}}_{ixI}^{(I)X}$,
$\tilde{\mathcal{E}}_{iyI}^{(I)X}$, 
$\tilde{\mathcal{E}}_{izI}^{(I)X}$, 
$\tilde{\mathcal{E}}_{ixI}^{(I)Y}$, 
$\tilde{\mathcal{E}}_{iyI}^{(I)Y}$, 
$\tilde{\mathcal{E}}_{izI}^{(I)Y}$, $\tilde{\Phi}_{iI}$.

\vspace{0.5em}

Similarly, the following data grids are to be computed at the surfaces $S,R,T$ to generate the complete field information at those surfaces:

\vspace{0.5em}

\textbf{At the surface \textit{S}:}

\vspace{0.5em}

$\tilde{k}_{ixS}^{(S)}$, 
$\tilde{k}_{iyS}^{(S)}$, 
$\tilde{k}_{izS}^{(S)}$, 
$\tilde{\mathcal{E}}_{ixS}^{(S)X}$,
$\tilde{\mathcal{E}}_{iyS}^{(S)X}$, 
$\tilde{\mathcal{E}}_{izS}^{(S)X}$, 
$\tilde{\mathcal{E}}_{ixS}^{(S)Y}$, 
$\tilde{\mathcal{E}}_{iyS}^{(S)Y}$, 
$\tilde{\mathcal{E}}_{izS}^{(S)Y}$, $\tilde{\Phi}_{iS}$;

$\tilde{k}_{rxS}^{(S)}$, 
$\tilde{k}_{ryS}^{(S)}$, 
$\tilde{k}_{rzS}^{(S)}$, 
$\tilde{\mathcal{E}}_{rxS}^{(S)X}$,
$\tilde{\mathcal{E}}_{ryS}^{(S)X}$, 
$\tilde{\mathcal{E}}_{rzS}^{(S)X}$, 
$\tilde{\mathcal{E}}_{rxS}^{(S)Y}$, 
$\tilde{\mathcal{E}}_{ryS}^{(S)Y}$, 
$\tilde{\mathcal{E}}_{rzS}^{(S)Y}$, $\tilde{\Phi}_{rS}$;

$\tilde{k}_{txS}^{(S)}$, 
$\tilde{k}_{tyS}^{(S)}$, 
$\tilde{k}_{tzS}^{(S)}$, 
$\tilde{\mathcal{E}}_{txS}^{(S)X}$,
$\tilde{\mathcal{E}}_{tyS}^{(S)X}$, 
$\tilde{\mathcal{E}}_{tzS}^{(S)X}$, 
$\tilde{\mathcal{E}}_{txS}^{(S)Y}$, 
$\tilde{\mathcal{E}}_{tyS}^{(S)Y}$, 
$\tilde{\mathcal{E}}_{tzS}^{(S)Y}$, $\tilde{\Phi}_{tS}$;

\vspace{0.5em}

\textbf{At the surface \textit{R}:}

\vspace{0.5em}

$\tilde{k}_{rxR}^{(R)}$, 
$\tilde{k}_{ryR}^{(R)}$, 
$\tilde{k}_{rzR}^{(R)}$, 
$\tilde{\mathcal{E}}_{rxR}^{(R)X}$,
$\tilde{\mathcal{E}}_{ryR}^{(R)X}$, 
$\tilde{\mathcal{E}}_{rzR}^{(R)X}$, 
$\tilde{\mathcal{E}}_{rxR}^{(R)Y}$, 
$\tilde{\mathcal{E}}_{ryR}^{(R)Y}$, 
$\tilde{\mathcal{E}}_{rzR}^{(R)Y}$, $\tilde{\Phi}_{rR}$;

\vspace{0.5em}

\textbf{At the surface \textit{T}:}

\vspace{0.5em}

$\tilde{k}_{txT}^{(T)}$, 
$\tilde{k}_{tyT}^{(T)}$, 
$\tilde{k}_{tzT}^{(T)}$, 
$\tilde{\mathcal{E}}_{txT}^{(T)X}$,
$\tilde{\mathcal{E}}_{tyT}^{(T)X}$, 
$\tilde{\mathcal{E}}_{tzT}^{(T)X}$, 
$\tilde{\mathcal{E}}_{txT}^{(T)Y}$, 
$\tilde{\mathcal{E}}_{tyT}^{(T)Y}$, 
$\tilde{\mathcal{E}}_{tzT}^{(T)Y}$, $\tilde{\Phi}_{tT}$.

\vspace{0.5em}

Here, the $\tilde{\Phi}_{jC}$ terms are the complete path-dependent phase functions of the fields $\tilde{\mathbf{E}}_{jC}^{(C)}$, interpreted similarly as $\tilde{\Phi}_{iI}$ [Eq. (\ref{Phi_iI})]. 
These path-dependent phases do not involve any sign-compensation convention for the field amplitude values [Appendix \ref{App_Amplitude}]; and hence we get $\tilde{\Phi}_{iS} = \tilde{\Phi}_{rS} = \tilde{\Phi}_{tS}$ --- a generalized phase-matching for non-planar waves.
The above data grids at all the $C$ surfaces, along with the phase-difference term $\Phi_E$ [Eq. (\ref{E0I})] and the reference phase $\Phi_O = 0$ [Eq. (\ref{EiII_WSE_full})], 
completely determine all the field profiles $\tilde{\mathbf{E}}_{jC}^{(C)}$.

The principle of energy conservation is utilized to verify the correctness of our simulation. We numerically verify all energy conservation relations of the following forms, and the ones derivable from these, by using the obtained field data:
\begin{subequations}
\begin{eqnarray}
& P_{0}^{X,Y} = \tilde{P}_{iI}^{X,Y} = \tilde{P}_{iS}^{X,Y}; & \label{P0=PiI} \\
& \tilde{P}_{rS}^{X,Y} = \tilde{P}_{rR}^{X,Y}; \hspace{1em} \tilde{P}_{tS}^{X,Y} = \tilde{P}_{tT}^{X,Y}; &\\
& \tilde{P}_{iS}^{X,Y} = \tilde{P}_{rS}^{X,Y} + \tilde{P}_{tS}^{X,Y}; \hspace{1em} \tilde{P}_{jC}^X + \tilde{P}_{jC}^Y = \tilde{P}_{jC} \,; &
\end{eqnarray}
\end{subequations}
where, $\tilde{P}_{jC}$ denotes the total power of the field $\tilde{\mathbf{E}}_{jC}^{(C)}$.
Moreover, at each point $\left( x^{(S)},y^{(S)},0^{(S)} \right)$, we numerically verify the intensity relations
\begin{eqnarray}
& \mathcal{I}_{iS}^X = \mathcal{I}_{rS}^X + \mathcal{I}_{tS}^X; \hspace{0.4em} 
\mathcal{I}_{iS}^Y = \mathcal{I}_{rS}^Y + \mathcal{I}_{tS}^Y; \hspace{0.4em} 
\mathcal{I}_{iS} = \mathcal{I}_{rS} + \mathcal{I}_{tS};\hspace{1.6em}  &
\end{eqnarray}
which are re-expressions of the well-known Fresnel formalism result: $\mbox{reflectivity} + \mbox{transmissivity} = 1$ \cite{SalehTeich, BornWolf, Jackson}.
These observations further verify the correctness of our simulation.

\subsection{Demonstration : Simulated Field Profiles}\label{Sub_SimProfiles}

\begin{figure*}[t]
\begin{center}
\includegraphics[width = \linewidth]{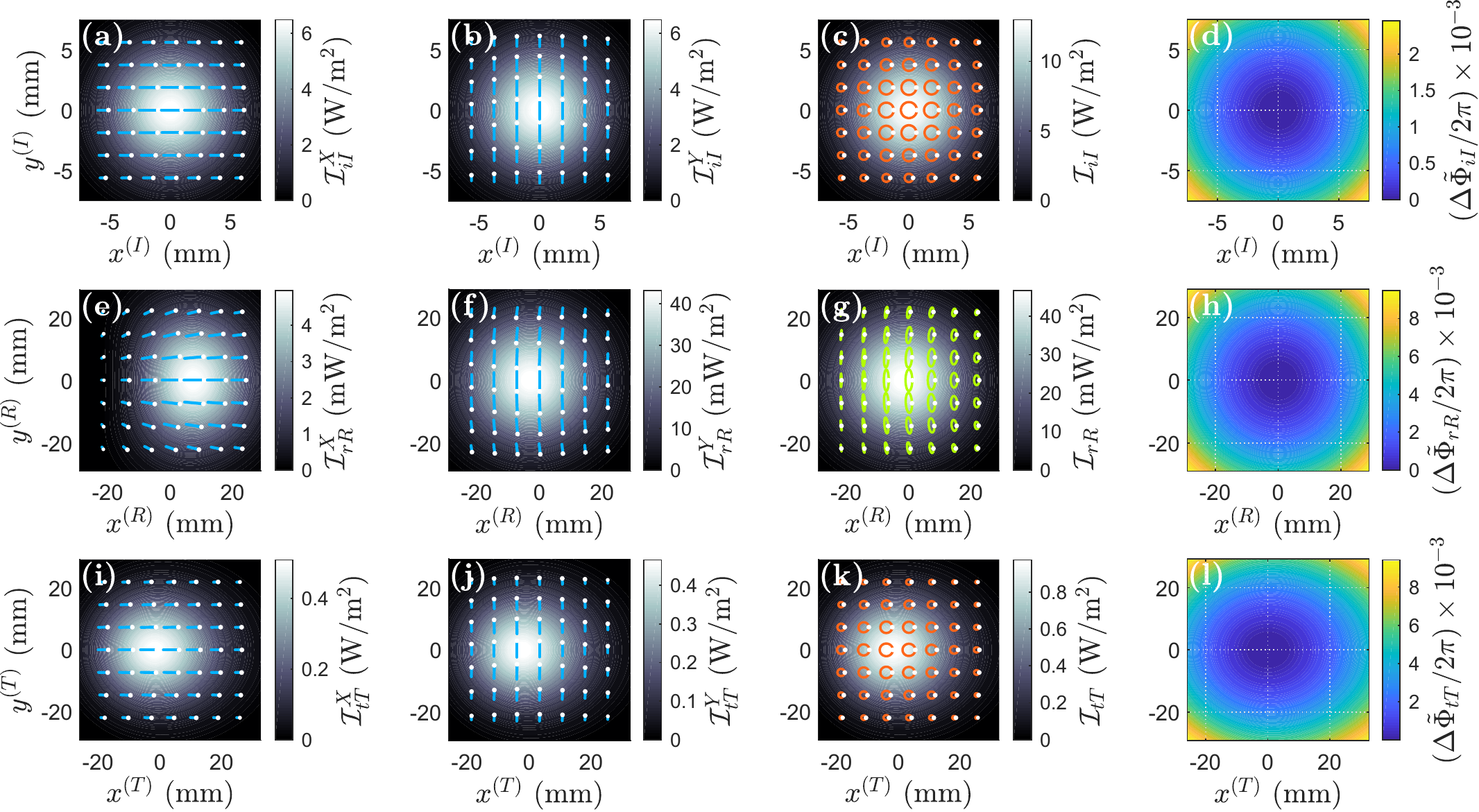}
\end{center}
\caption{\small{
Graphical representations of the electric fields $\tilde{ \mathbf{E}}_{jJ}^{(J)}$ in terms of the quantities $\tilde{\boldsymbol{ \mathcal{E}}}_{jJ}^{(J)X}$, $\tilde{\boldsymbol{ \mathcal{E}}}_{jJ}^{(J)Y}$, $\tilde{\boldsymbol{ \mathcal{E}}}_{jJ}^{(J)}$ and $\Delta\tilde{\Phi}_{jJ}$; as functions of coordinates $\left( x^{(J)}, y^{(J)} \right)$ ($J = I, R, T$); for the chosen simulation parameters (full description in Subsection \ref{Sub_SimProfiles}).
The profiles are for 
\textbf{(a)} $\tilde{\boldsymbol{ \mathcal{E}}}_{iI}^{(I)X}$; 
\textbf{(b)} $\tilde{\boldsymbol{ \mathcal{E}}}_{iI}^{(I)Y}$; 
\textbf{(c)} $\tilde{\boldsymbol{ \mathcal{E}}}_{iI}^{(I)}$; 
\textbf{(d)} $\Delta \tilde{\Phi}_{iI}$; 
\textbf{(e)} $\tilde{\boldsymbol{ \mathcal{E}}}_{rR}^{(R)X}$; 
\textbf{(f)} $\tilde{\boldsymbol{ \mathcal{E}}}_{rR}^{(R)Y}$; 
\textbf{(g)} $\tilde{\boldsymbol{ \mathcal{E}}}_{rR}^{(R)}$; 
\textbf{(h)} $\Delta \tilde{\Phi}_{rR}$; 
\textbf{(i)} $\tilde{\boldsymbol{ \mathcal{E}}}_{tT}^{(T)X}$; 
\textbf{(j)} $\tilde{\boldsymbol{ \mathcal{E}}}_{tT}^{(T)Y}$; 
\textbf{(k)} $\tilde{\boldsymbol{ \mathcal{E}}}_{tT}^{(T)}$; 
\textbf{(l)} $\Delta \tilde{\Phi}_{tT}$.
}}
{\color{grey}{\rule{\linewidth}{1pt}}}
\label{fig_field_profiles}
\end{figure*}

In this subsection we give an example of field-data generation by taking some specific simulation-parameter values. We take $\lambda = 632.8$ nm; $n_1 = 1$, $n_2 = 1.52$; input power $P_0 = 1$ mW (this determines $\mathcal{E}_{00}$ in Eqs. (\ref{E0xI}, \ref{E0yI}) ); $\theta_E = 45^\circ$; $\Phi_E = \pi/2$; $w_0 = 7$ mm; $f = 3.5$ cm (hence, $2\theta_D \approx 22.62^\circ$); 
$\theta_{i0} = 45^\circ$; distances $O_I O_S = O_S O_{J'} = 5$ cm. 
We have chosen a large enough divergence $2\theta_D$ to demonstrate that our formalism is applicable --- among other optical waves --- to non-paraxial beams as well; as opposed to the standard formalisms, such as that of Ref. \cite{BARev}, which are applicable only to paraxial beams.

We represent the computationally generated field data by using the specially designed profiles of Fig. \ref{fig_field_profiles}. To explain these profiles, we first consider the projection of the field $\tilde{\boldsymbol{ \mathcal{E}}}_{iI}^{(I)X}$ on the surface $I$, given by
\begin{equation}
\tilde{\boldsymbol{ \mathcal{E}}}_{iI}^{\prime (I) X} = \tilde{\mathcal{E}}_{ixI}^{(I)X} \hat{\mathbf{x}}^{(I)} + \tilde{\mathcal{E}}_{iyI}^{(I)X} \hat{\mathbf{y}}^{(I)};
\end{equation}
(a detailed analysis of such a projection, including its geometric phase characteristics, is discussed in Appendix \ref{App_GeoPhase}). We then consider an auxiliary field 
\begin{equation}
\tilde{\mathbf{E}}_{iI}^{\prime (I) X} = \tilde{\boldsymbol{ \mathcal{E}}}_{iI}^{\prime (I) X} e^{i \omega t};
\end{equation}
whose real part gives the physically appearing field projection at the surface $I$. If this real field at any time $t$ and at any point $\left( x^{(I)}, y^{(I)} \right)$ is represented by an arrow, then the variation of $\omega t$ in range $[0,2\pi)$ gives the complete trajectory of the tip of this arrow; which, in the case of $\tilde{\mathbf{E}}_{iI}^{\prime (I) X}$, is a line-segment --- since this field is locally linearly polarized. In Fig. \ref{fig_field_profiles}(a), the field amplitude vector profile $\tilde{\boldsymbol{ \mathcal{E}}}_{iI}^{(I) X}$ is represented by such line-segments at the surface $I$. The white dot associated with each line-segment represents the arrow-tip position of $\mathfrak{Re} \left( \tilde{\mathbf{E}}_{iI}^{\prime (I) X} \right)$ at $t = 0$. Additionally, the intensity profile $\mathcal{I}_{iI}^{X}$ of the field $\tilde{\boldsymbol{ \mathcal{E}}}_{iI}^{(I) X}$ is shown in the background.

A similar representation of the field profile $\tilde{\boldsymbol{ \mathcal{E}}}_{iI}^{(I) Y}$, involving the projection and auxiliary fields
\begin{equation}
\tilde{\boldsymbol{ \mathcal{E}}}_{iI}^{\prime (I) Y} = \tilde{\mathcal{E}}_{ixI}^{(I)Y} \hat{\mathbf{x}}^{(I)} + \tilde{\mathcal{E}}_{iyI}^{(I)Y} \hat{\mathbf{y}}^{(I)},
\hspace{1em}
\tilde{\mathbf{E}}_{iI}^{\prime (I) Y} = \tilde{\boldsymbol{ \mathcal{E}}}_{iI}^{\prime (I) Y} e^{i \omega t},
\end{equation}
and the intensity profile $\mathcal{I}_{iI}^{Y}$, is given in Fig. \ref{fig_field_profiles}(b).

The projection of the field $\tilde{\boldsymbol{ \mathcal{E}}}_{iI}^{(I)}$ has a complex form
\begin{eqnarray}
&& \tilde{\boldsymbol{ \mathcal{E}}}_{iI}^{\prime (I)} = \left( \tilde{\mathcal{E}}_{ixI}^{(I)X} + e^{i\Phi_E} \tilde{\mathcal{E}}_{ixI}^{(I)Y} \right) \hat{\mathbf{x}}^{(I)} \nonumber\\ 
&& \hspace{7.5em} + \left( \tilde{\mathcal{E}}_{iyI}^{(I)X} + e^{i\Phi_E} \tilde{\mathcal{E}}_{iyI}^{(I)Y} \right) \hat{\mathbf{y}}^{(I)}.
\end{eqnarray}
So, the representative arrow-tip of the corresponding real-auxiliary field $\mathfrak{Re} \left( \tilde{\mathbf{E}}_{iI}^{\prime (I)} \right)$ traverses an elliptical trajectory with the variation of $\omega t$. In Fig. \ref{fig_field_profiles}(c), the profile $\tilde{\boldsymbol{ \mathcal{E}}}_{iI}^{(I)}$ is represented by such ellipses (dark orange: right elliptical polarization; light green: left elliptical polarization) at the surface $I$; with the white dot associated with each ellipse representing the arrow-tip position of $\mathfrak{Re} \left( \tilde{\mathbf{E}}_{iI}^{\prime (I)} \right)$ at $t = 0$. The corresponding intensity profile $\mathcal{I}_{iI}$ is shown in the background. One can visualize that the field profile of Fig. \ref{fig_field_profiles}(b) is first given an additional phase $\Phi_E$; and then is superposed with the field profile of Fig. \ref{fig_field_profiles}(a) to obtain the profile of Fig. \ref{fig_field_profiles}(c).

Figures \ref{fig_field_profiles}(a--c) do not contain any contribution from the phase function $\tilde{\Phi}_{iI}$, which is separately represented in Fig. \ref{fig_field_profiles}(d). The functional form of $\tilde{\Phi}_{iI}$ is given simply by Eq. (\ref{Phi_iI}). The $\tilde{\Phi}_{iI}$ value at the origin $O_I$ is $\Phi_{OI}^0 = n_1 k \,|f|$. Using this value as a reference, in Fig. \ref{fig_field_profiles}(d) we plot a relative phase function 
\begin{equation}
\Delta\tilde{\Phi}_{iI} = \tilde{\Phi}_{iI} - \Phi_{OI}^0 \, ; \label{DeltaPhI=PhI-PhOI}
\end{equation}
which well-represents the phase function $\tilde{\Phi}_{iI}$. As understood from the discussion of Subsection \ref{Sub_WSE}, all surface elements corresponding to the same wavefront have the same phase; whereas, surface elements corresponding to different wavefronts have different phases [Fig. \ref{fig_WSEComps}]. This phase variation over the surface $I$ is represented by the profile of Fig. \ref{fig_field_profiles}(d). In other words, the phase variation of Fig. \ref{fig_field_profiles}(d) signifies the wavefront-curvature of the field $\tilde{\mathbf{E}}_{iI}^{(I)}$. As per our constructed model, the 2D gradient of $\tilde{\Phi}_{iI}$ at the surface $I$ is given by $\tilde{k}_{ixI}^{(I)}  \, \hat{\mathbf{x}}^{(I)} + \tilde{k}_{iyI}^{(I)} \, \hat{\mathbf{y}}^{(I)}$, which we have verified computationally.

Thus, Figs. \ref{fig_field_profiles}(a--d) give a full representation of the complete incident field $\tilde{\mathbf{E}}_{iI}^{(I)}$ of Eq. (\ref{EiII_WSE_full}). In a similar way, Figs. \ref{fig_field_profiles}(e--h) represent the complete reflected field $\tilde{\mathbf{E}}_{rR}^{(R)}$ and Figs. \ref{fig_field_profiles}(i--l) represent the complete transmitted field $\tilde{\mathbf{E}}_{tT}^{(T)}$ [Subsection \ref{Sub_Recombination}] by involving the corresponding projection fields, auxiliary fields and intensity profiles. The relative phase functions shown in Figs. \ref{fig_field_profiles}(h, l) are $\Delta\tilde{\Phi}_{j'J'} = \tilde{\Phi}_{j'J'} - \Phi_{OJ'}^0$, where the reference phase term $\Phi_{OJ'}^0$ is the phase of the field $\tilde{\mathbf{E}}_{j'J'}^{(J')}$ at the origin $O_{J'}$. We have also obtained similar representative profiles for the complete fields $\tilde{\mathbf{E}}_{jS}^{(S)}$ in the simulation.



\vspace{1em}

\textbf{Inhomogeneous Polarization and Beam Shifts:}
Figures \ref{fig_field_profiles}(a--c, e--g, i--k) show that the field profiles $\tilde{\boldsymbol{\mathcal{E}}}_{j'J'}^{(J')X}$, $\tilde{\boldsymbol{\mathcal{E}}}_{j'J'}^{(J')Y}$, $\tilde{\boldsymbol{\mathcal{E}}}_{j'J'}^{(J')}$ are radically different from the corresponding $\tilde{\boldsymbol{\mathcal{E}}}_{iI}^{(I)X}$, $\tilde{\boldsymbol{\mathcal{E}}}_{iI}^{(I)Y}$, $\tilde{\boldsymbol{\mathcal{E}}}_{iI}^{(I)}$ incident-field profiles (i.e. the Fig.-\ref{fig_field_profiles}(e, i)-profiles are different from the corresponding Fig.-\ref{fig_field_profiles}(a)-profile; the Fig.-\ref{fig_field_profiles}(f, j)-profiles are different from the corresponding Fig.-\ref{fig_field_profiles}(b)-profile; the Fig.-\ref{fig_field_profiles}(g, k)-profiles are different from the corresponding Fig.-\ref{fig_field_profiles}(c)-profile). The spatial variations of the $\tilde{\boldsymbol{\mathcal{E}}}_{j'J'}^{(J')X}$, $\tilde{\boldsymbol{\mathcal{E}}}_{j'J'}^{(J')Y}$, $\tilde{\boldsymbol{\mathcal{E}}}_{j'J'}^{(J')}$ polarization-projections are not simply the distortion-effect due to the wavefront-curvature [Appendix \ref{App_GeoPhase}] (which, nevertheless, contributes as well); but the momentum-spatial variation of the $\tilde{\mathbf{j'}}_S^{(S)}$ matrices [Eqs. (\ref{rMat}, \ref{tMat})] via Eqs. (\ref{Aterms}), for the different constituent plane waves, introduces a fundamental inhomogeneity in the polarization profiles. This is the same underlying physical phenomenon which is described as the spatial-dispersion effect by Bliokh and Aiello \cite{BARev}. In addition, the transmitted wavefront-shape itself is distorted as compared to the incident wavefront-shape [Appendix \ref{App_gTS}], as understood from Fig. \ref{fig_field_profiles}(l).

These polarization variations are coupled with variations in intensity profiles, as observed in Fig. \ref{fig_field_profiles}. Each $\mathcal{I}_{j'J'}$ intensity profile ($J' = R,T$) is considerably different from the $\mathcal{I}_{iI}$ profile because of the above effects. While the $\mathcal{I}_{iI}$ centroid appears at the origin $O_I$ in our simulated model, the $\mathcal{I}_{j'J'}$ centroid in general appears at a point $\left( x_G^{(J')}, y_F^{(J')} \right)$ --- shifted from the origin $O_{J'}$. The longitudinal shift $x_G^{(J')}$ is a manifestation of GH shift \cite{GH, Artmann, RaJW, AntarYM, McGuirk, ChanCC, Porras, AielloArXiv, AielloArXiv2, Aiello2008, Aiello2009, Qin2011, BARev}; and the transverse shift $y_F^{(J')}$ is a manifestation of IF shift \cite{Fedorov, Schilling, Imbert, Player, FVG, Liberman, Onoda, Bliokh2006, Bliokh2007, HostenKwiat, AielloArXiv2, Aiello2008, Aiello2009, Qin2011, BARev}. In the simulation, we can vary the distance $O_S O_{J'}$ [Fig. \ref{fig_setup}]; and create plots of $x_G^{(J')}$ and $y_F^{(J')}$ as functions of $O_S O_{J'}$. Such plots contain sufficient information to determine the spatial and angular GH and IF shifts \cite{BARev} for both the reflected and transmitted beams at the dielectric interface $S$.

We have observed in the simulation that no transverse shifts occur to the individual $\mathcal{I}_{j'J'}^X$ and $\mathcal{I}_{j'J'}^Y$ profiles 
(i.e. the centroids of the Fig.-\ref{fig_field_profiles}(e, i)-profiles are not transverse-shifted with respect to the centroid of the Fig.-\ref{fig_field_profiles}(a)-profile; and the centroids of the Fig.-\ref{fig_field_profiles}(f, j)-profiles are not transverse-shifted with respect to the centroid of the Fig.-\ref{fig_field_profiles}(b)-profile).
They undergo only longitudinal shifts, say $x_{G}^{(J')X}$ and $x_{G}^{(J')Y}$, which are considerably different from each other. Since $\tilde{\boldsymbol{\mathcal{E}}}_{j'J'}^{(J')X}$ is mostly $\hat{\mathbf{x}}^{(J')}$-polarized [Figs. \ref{fig_field_profiles}(e, i)] and $\tilde{\boldsymbol{\mathcal{E}}}_{j'J'}^{(J')Y}$ is mostly $\hat{\mathbf{y}}^{(J')}$-polarized [Figs. \ref{fig_field_profiles}(f, j)], the difference between $x_{G}^{(J')X}$ and $x_{G}^{(J')Y}$ appears as an effective polarization separation. However, no polarization-induced wavevector-separation (i.e. birefringence) occurs in the system; and the effective polarization separation occurs simply because of the difference between the $\mathcal{I}_{j'J'}^X$ and $\mathcal{I}_{j'J'}^Y$ profiles. We refer to this phenomenon as a pseudo-birefringence at the dielectric interface, which can be easily observed experimentally by implementing optical weak-measurement--post-selection methods \cite{AAV, DSS, RSH, Dennis, Gotte, HostenKwiat, Aiello2008}.

The projected field $\tilde{\boldsymbol{\mathcal{E}}}_{j'J'}^{\prime (J')}$ can be decomposed into $\hat{\boldsymbol{\sigma}}^{\pm}$ spin-polarization states. If the centroid positions of these spin-decomposed fields are $\left( x_\pm^{(J')}, y_\pm^{(J')} \right)$, then $x_\pm^{(J')}$ signify longitudinal spin shifts \cite{Qin2011} and $y_\pm^{(J')}$ signify transverse spin shifts (which include spin-Hall shifts) 
\cite{Liberman, Onoda, Bliokh2006, Bliokh2007, HostenKwiat, Qin2011, XieSHELinIF} --- both of which can be determined in our simulation. Our simulated system thus shows that these spin-shifts appear not due to any wavevector deflection, but due to the difference between the spin-decomposed field intensity profiles.

Finally, all the effects mentioned in this subsection rely on beam-divergence. The extent of these effects can be directly controlled by varying the divergence angle $2\theta_D$ of the incident beam-field $\tilde{\mathbf{E}}_{iI}^{(I)}$. In particular, the variation of these effects, as the $\tilde{\mathbf{E}}_{iI}^{(I)}$ beam-field is transformed from paraxial to non-paraxial domain, can be effectively studied by using our simulation. Our works on these beam-shifts and other relevant effects will be reported elsewhere (e.g. Ref. \cite{CLEO2020}).


\section{Applicability and Generalization}\label{Sec_ApplGen}

Our formalism not only re-expresses the dielectric reflection-transmission problem of complex optical waves 
in an elegant mathematical structure; but also serves the purpose of a unified mathematical formalism that enables the analysis of all related problems in a single generalized method. The compact calculating steps also ensure a remarkable computational efficiency [Section \ref{Sec_CompAspects}] while simulating a reflection-transmission model of complex optical waves at a plane dielectric interface.

We have demonstrated our formalism in terms of partial reflection-transmission of a model optical wave at a single plane dielectric interface. 
However, because of the generic nature of the $\tilde{\mathbf{j'}}_S^{(S)}$ matrices [Eq. (\ref{rMat}, \ref{tMat})], our formalism is applicable to a significantly generalized class of optical systems. Here we describe some direct applications of our formalism, followed by generalizations.

\subsection{Direct Applications}

Taking the $\tilde{\mathbf{E}}_{iI}^{(I)}$ field as a paraxial Gaussian beam field, our formalism explores the properties of the corresponding reflected and transmitted fields $\tilde{\mathbf{E}}_{j'J'}^{(J')}$, including their GH and IF shifts --- thus confirming the results of Dennis, G\"otte, L\"offler \cite{Dennis, Gotte, GotteLofflerDennis} and Bliokh-Aiello \cite{BARev}. Expressing $\tilde{\mathbf{E}}_{j'J'}^{(J')}$ in terms of the $\hat{\boldsymbol{\sigma}}^\pm$ spin-polarization eigenstates, the formalism reproduces the SHEL results obtained by Hosten-Kwiat \cite{HostenKwiat} and Xie et al. \cite{XieSHELinIF}.
Taking $\tilde{\mathbf{E}}_{iI}^{(I)}$ as a dipole radiation field \cite{Jackson}, our formalism reproduces Berry's results on reflected dipole radiation \cite{Berry435}.
Thus, our single formalism can analyse systems involving optical waves ranging from collimated and paraxial beams to full spherical waves --- including non-paraxial beams and other intermediate waves with arbitrary divergence/convergence. 
One such example is the system described by Barczyk et al. \cite{VortexBrewster}, which shows the generation of vectorial vortex due to Brewster reflection of a non-vortex beam. We have explored the generic presence of both phase and polarization singularities in a Brewster-reflected paraxial beam by using this formalism \cite{CLEO2020}.

The Fourier decomposition case enables us to analyse a very large class of complex optical waves. As a significant example, we mention here the reflection-transmission problem of vortex beams. Li et al. have explained how to express a vortex beam field in terms of Fourier component plane waves \cite{LiVortexRT}. By using these component waves in our formalism, we can find the reflected and transmitted vortex beam fields. Significant results in the literature, such as the ones by Dennis-Götte \cite{DennisGotteVortex} and Yavorsky-Brasselet \cite{YavorskyBrasselet}, can thus be efficiently reproduced by using our formalism. Subsequently, we are also able to search for new and significant effects in complex optical systems by using the corresponding simulated models.




\subsection{Total Internal Reflection}


For $n_1 > n_2$ and $\theta_i > \theta_c$, the critical angle \cite{SalehTeich, BornWolf, Jackson}, we get $ \left( n_1^2 k^2 - k_{izS}^{(S)\;2} \right) > n_2^2 k^2 $ [Eq. (\ref{kjzSS})], which makes the component $k_{tzS}^{(S)}$ purely imaginary. This is the case of total internal reflection.
Equations (\ref{EiII_WSE}--\ref{EtTT}) are applicable to the total internal reflection case also; however, the terms $\mathbf{k}_{tT}^{(S)}$, $\Phi_{TS}$ and $g_{TS}$ [Eq. (\ref{EtTT})] require special interpretations here.

We know that, $\mathbf{k}_{tT}^{(S)} = \mathbf{k}_{tS}^{(S)}$, since the wavevector remains unchanged due to propagation. 
For imaginary $k_{tzS}^{(S)}$, we write $k_{tzS}^{(S)} = i\kappa_{tzS}^{(S)}$, where $\kappa_{tzS}^{(S)} = \left| k_{tzS}^{(S)} \right|$. Then, the wavevector $\mathbf{k}_{tS}^{(S)}$ is expressed as
\begin{equation}
\mathbf{k}_{tS}^{(S)} = k_{txS}^{(S)}\, \hat{\mathbf{x}}^{(S)} + k_{tyS}^{(S)}\, \hat{\mathbf{y}}^{(S)} + i\kappa_{tzS}^{(S)}\, \hat{\mathbf{z}}^{(S)}.
\end{equation}
The phase term $\Phi_{TS}$ is then obtained in the form
\begin{equation}
\Phi_{TS} = \mathbf{k}_{tS}^{(S)} \cdot \Delta \mathbf{r}_{TS}^{(S)} = \Phi_{xyTS} + i\Phi_{zTS};\label{PhiTS_TIR}
\end{equation}
\vspace{-1.5em}
\begin{eqnarray}
\hspace{-2em}\mbox{where,} \hspace{1em} \Phi_{xyTS} &=& k_{txS}^{(S)}\, \Delta x^{(S)} + k_{tyS}^{(S)}\, \Delta y^{(S)};\\
\Phi_{zTS} &=& \kappa_{tzS}^{(S)}\, z^{(S)}.
\end{eqnarray}
Using Eq. (\ref{PhiTS_TIR}) in Eq. (\ref{EtTT}), we get
\begin{equation}
\mathbf{E}_{tT}^{(T)} 
= g_{TS}\, g_{SI} \tilde{\mathbf{R}}_{TS}\, \tilde{\mathbf{t}}_{S}^{(S)} \tilde{\mathbf{R}}_{SI} \mathbf{E}_{iI}^{(I)} e^{-\Phi_{zTS}}e^{i \left(\Phi_{xyTS} + \Phi_{SI}\right)}. \label{EtTT_TIR}
\end{equation}
So, the phase term $\Phi_{xyTS}$ implies a transmitted wave propagation on the surface $S$ along the direction $k_{ixS}^{(S)}\, \hat{\mathbf{x}}^{(S)} + k_{iyS}^{(S)} \, \hat{\mathbf{y}}^{(S)}$, while the factor $e^{-\Phi_{zTS}}$ gives an exponential decay of the field amplitude along $\hat{\mathbf{z}}^{(S)}$ \cite{BornWolf, Jackson}.

The factor $g_{TS}$ is unity for a Fourier component plane wave. However, the wavefront for the surface-element decomposition case is to be reinterpreted here. We can define the wavefront as the surface in 3D space on which the electric field $\mathbf{E}_{tT}^{(T)}$ exists. A surface element on this wavefront expands/contracts as the wave propagates on the interface. This change in size determines the factor $g_{TS}$ in the wavefront-surface-element decomposition case.

With the above reinterpretations, our formalism is readily applicable to the problems involving total internal reflection of complex optical waves.

\section{Computational Aspects}\label{Sec_CompAspects}

A standard and widely-used simulation method for electromagnetic field problems is the finite difference time domain (FDTD) method \cite{FDTDBook}, which numerically solves differential equations to compute electromagnetic field data. While this method requires heavy computation, it has immense potential to be used in complex systems where analytical solutions are not possible. However, the reflection/transmission problem of complex optical waves at a plane dielectric interface is an analytically solvable problem. To implement the FDTD method in such an analytically solvable problem is practically a waste of computational time. Our formalism has both analytical [Sections \ref{Sec_Derivation}, \ref{Sec_Formalism}] and computational [Section \ref{Sec_Simulation}] aspects; and it works based on the idea of using analytical expressions wherever applicable. Our algorithm, as one can qualitatively visualize, first creates data-points; and then simply evaluates already-derived analytical expressions there --- resulting in the first stage of high computational efficiency-gain over the standard FDTD method. 

The second stage of efficiency-gain arises from the fact that the FDTD simulation of a 3D system usually requires a 3D mesh. However, our algorithm requires only 2D data grids at the $C$ surfaces even without loosing any information of the optical fields --- thus reducing the problem to a 2D simulation. This results in the second stage of high efficiency-gain.

The third stage in efficiency-gain results from the use of the $\tilde{\mathbf{j'}}_S^{(S)}$ matrices [Eqs. (\ref{rMat}, \ref{tMat})]. As explained in Subsection \ref{Sub_AltDerivation}, by directly using the $\tilde{\mathbf{j'}}_S^{(S)}$ matrices we avoid all intermediate calculating steps involving the local $S'$ coordinate systems. Avoiding all the corresponding intermediate computations in the simulation thus gives a remarkable efficiency-gain as compared to any simulation which uses Fresnel coefficients.

Our algorithm also provides a remarkable gain in numerical accuracy. Since the computation is based on evaluation of analytical expressions, the data are remarkably accurate as compared to the FDTD simulation data which are generated by numerical solutions of differential equations. The only sources of numerical errors are the numerical integrations, which are used while finding the centroid positions of the intensity profiles [Subsection \ref{Sub_SimProfiles}]. However, given the above high efficiency-gains, we can use a much larger size of data-grids at the $C$ surfaces, as compared to the number of surface-points in a corresponding FDTD simulation that would take a comparable computational time in the same hardware. Hence, we achieve a remarkable gain in numerical accuracy as well.

Apart from the efficiency and accuracy gains, we are also able to make a choice between the Fourier decomposition and the wavefront-surface-element decomposition methods. As discussed in Section \ref{Sec_Formalism}, the wavefront-surface-element decomposition is a straightforward method, easily implemented on spherical, cylindrical and other simple-shaped wavefronts. On the other hand, the Fourier decomposition is a more general approach --- the preferred method for general optical waves where a simple surface-element decomposition is not achieved.


We have written our current simulation codes in Matlab (R2018a). We have run these codes in a Dell G7 7588 laptop that houses an Intel 8th Gen Core i9-8950HK CPU, 16 GB RAM and an NVIDIA GeForce GTX 1060 GPU. We have implemented GPU-computation wherever applicable. 
In a sample run we use data-grids of size $1000 \times 1000$. During the simulation-run, the CPU operates at a boosted clock-speed $\sim 4.2$ GHz; while Matlab uses $\sim 1.4$ GB RAM and $\sim 200$ MB GPU memory. The time taken by this run, computing all the functional data-grids mentioned in Subsection \ref{Sub_DataGrids}, is $\sim 20$ s.


\section{Conclusion}

We have introduced a generalized formalism, by which the reflection and transmission of complex optical waves at a plane dielectric interface 
can be analysed completely with remarkable computational efficiency. The central step in this formalism is the application of two generalized matrix operators --- the reflection and transmission coefficient matrices --- to the constituent plane-wave fields of the considered incident complex wave to obtain the corresponding reflected and transmitted constituent plane-wave fields. This step acts as a physically equivalent, but mathematically elegant and computationally efficient replacement to the usual Fresnel formalism. We have derived these matrices; and 
have given the complete mathematical details of a physical scenario by implementing these matrices --- 
thus describing our generalized matrix transformation formalism. The use of exact 3D wavevector and electric field expressions gives the  formalism a very generic nature, which enables us to analyse a very large class of complex optical waves, with automatically retaining the wavefront curvature 
and geometric phase information.

We have demonstrated the working of our formalism by using it in a simulated Gaussian beam model. We have demonstrated how we can analytically create the incident field information; computationally generate the reflected and transmitted field data; and extract physically significant information about the optical fields from the generated data. 
While our present simulated model utilizes wavefront-surface-element decomposition method, we are also able to utilize Fourier decomposition method in the same formalism for various complex waves in general.
In this way, simulated models can be created for a very large class of dielectric reflection-transmission problems of complex optical waves, including vortex beam problems, total internal reflection --- all realizable based on our single generalized matrix transformation formalism.  
We have briefly discussed how our formalism, under different conditions, reproduces many novel optical phenomena described in the current literature.


\section*{APPENDICES}

\appendix

\section{Electric Field Amplitude and Phase Conventions}\label{App_Amplitude}

From Fresnel coefficient applications \cite{Jackson, SalehTeich, BornWolf} with $n_2 > n_1$, we know that the TM and TE reflected field projections at the dielectric interface have directions opposite to the corresponding incident field projections, for the angle-of-incidence ranges $0 \leq \theta_i < \theta_B$ (Brewster angle) and $0 \leq \theta_i < 90^\circ$ respectively.
Thus, direction-flips of the concerned incident fields are observed upon reflection in these cases.
It is a usual practice to identify the unsigned electric field value as the amplitude; and to interpret the sign-flip as a phase change of $\pi$ upon reflection.

However, even if the above representation is suitable (but not compulsory) for uniformly polarized plane waves, it creates inconveniences for the case of inhomogeneously polarized fields --- the kind of fields we deal with in the present work. To demonstrate this with an example, we consider a plane-wave beam-field of the form
\begin{eqnarray}
& \mathbf{E} = \boldsymbol{\mathcal{E}} \, e^{i \, \left( \mathbf{k}\cdot\mathbf{r} \, - \, \omega t \, + \, \Phi_O \right)} = \mathcal{E} \hat{\mathbf{x}} \, e^{i \, \left( \mathbf{k}\cdot\mathbf{r} \, - \, \omega t \, + \, \Phi_O \right)}; & \label{Amp_E=E_1} \\  
& \mathcal{E} = \mathcal{E}_0 \left( \dfrac{x}{w_0} \right) e^{- \rho^{2}/w_0^2}; & \label{Amp_E=E_2}
\end{eqnarray}
where, the various terms have their usual meanings ($\Phi_O = $ reference phase term); and we have omitted the coordinate system identifier superscript `$(C)$' for simplicity, because all quantities are expressed in the same $C$ coordinate system. The function $\mathcal{E} \equiv \mathcal{E}(x,y)$ is a 2D Hermite-Gaussian function with range of values $\mathcal{E} < 0$ for $x < 0$; $\mathcal{E} \geq 0$ for $x \geq 0$. To use the non-negative amplitude convention in this case, one can write
\begin{eqnarray}
\mathcal{E} = \mathcal{E}' e^{i \Phi_\pi}; \hspace{1em} \mathcal{E}' = |\mathcal{E}|; \hspace{1em} 
\Phi_\pi = \left\{ \begin{array}{cl}
0 & \mbox{for } x \geq 0;\\
\pi & \mbox{for } x < 0;\\
\end{array}\right. \label{Amp_E=EpPhi}
\end{eqnarray}
and then re-express $\mathbf{E}$ as
\begin{equation}
\mathbf{E} = \mathcal{E}' \hat{\mathbf{x}} \, e^{i \, \left( \mathbf{k}\cdot\mathbf{r} \, - \, \omega t \, + \, \Phi_O \, + \, \Phi_\pi \right)}. \label{Amp_E=Ep}
\end{equation}
The function $\mathcal{E}'$ is then identified as the non-negative amplitude; while the sign-flip is compensated by the discontinuous phase function $\Phi_\pi$. The $\pi$-discontinuity in $\Phi_\pi$ at $x = 0$ [Eq. (\ref{Amp_E=EpPhi})] is manifested in the field-expression of Eq. (\ref{Amp_E=Ep}) as a $\lambda/2$ step in the wavefront. Thus, the non-negative amplitude convention re-expresses the plane-wave beam-field $\mathbf{E}$ [Eq. (\ref{Amp_E=E_1})] as a beam-field with wavefront-dislocation \cite{Gbur} [Eq. (\ref{Amp_E=Ep})]. 
While such a convention is not inconvenient for either only $\mathcal{E} \geq 0$ fields or only $\mathcal{E} < 0$ fields, it creates inconveniences for complicated fields with sign-flips --- by introducing $\lambda/2$ dislocations at the corresponding places on the wavefronts.

To resolve this problem, we discard the non-negative amplitude convention for general purposes. We consider $\mathcal{E}$ [Eq. (\ref{Amp_E=E_2})] as the amplitude value \cite{FOOTNOTE_AmpDirection}; and the vector $\boldsymbol{\mathcal{E}} = \mathcal{E} \hat{\mathbf{x}}$ [Eq. (\ref{Amp_E=E_1})] as the amplitude vector --- without interpreting sign-flips as $\pi$ phase jumps. In general, we consider vector functions of the form
\begin{eqnarray}
\hspace{-1em} \boldsymbol{\mathcal{E}} &=& \boldsymbol{\mathcal{E}}_1 + e^{i\Phi} \boldsymbol{\mathcal{E}}_2 \nonumber\\
&\equiv & (a_x + i b_x) \, \hat{\mathbf{x}} + (a_y + i b_y) \, \hat{\mathbf{y}} + (a_z + i b_z) \, \hat{\mathbf{z}} \label{Amp_Eform}
\end{eqnarray}
as the complex amplitude vector functions. Such a complex amplitude vector physically represents an appropriately oriented polarization ellipse in 3D space.
We consider the electric fields $\mathbf{E}_1 = \boldsymbol{\mathcal{E}} \, e^{i \, \left( \mathbf{k}\cdot\mathbf{r} \, - \, \omega t \right)}$ and $\mathbf{E}_2 = (-\boldsymbol{\mathcal{E}}) \, e^{i \, \left( \mathbf{k}\cdot\mathbf{r} \, - \, \omega t \, + \, \pi \right)}$ as two different fields for a general analysis; and bring in the context of their physical equality only if it is required for specific interpretations.
The amplitude vectors explicitly contain only the phase-difference term $\Phi$ as in Eq. (\ref{Amp_Eform}) (e.g. $\Phi_E$ of Eq. (\ref{E0I}) ). 
Additionally, the geometric phase characteristics of the fields are derived from these complex amplitude vector functions [Appendix \ref{App_GeoPhase}].

On the other hand, we also do not restrict the phase of a field within any limited range, e.g. $[-\pi/2, \pi/2]$ or $[0, \pi]$. The phase contains three terms [Eq. (\ref{Amp_E=E_1})] --- the path dependent $\mathbf{k} \cdot \mathbf{r}$ term; the $\omega t$ term; and the reference phase term $\Phi_O$, if any (e.g. as in Eq. (\ref{EiII_Fourier}) ) --- but no sign-compensating $\pi$-jump.

A closely related but different computational $\pi$-phase-jump situation arises while expressing the complex quantities $a_s + i \, b_s$ ($s = x, y, z$) of Eq. (\ref{Amp_Eform}) in the form
\begin{equation}
a_s + i \, b_s = \left( a_s^2 + b_s^2 \right)^{\frac{1}{2}} e^{i \tan^{-1}(b_s/a_s)}. \label{Amp_a+ib=rexp}
\end{equation}
While Eq. (\ref{Amp_a+ib=rexp}) is mathematically correct, a computer code simply takes the value of $\tan^{-1}(b_s/a_s)$ in the principal value range $[-\pi/2, \pi/2]$; and hence, a discontinuity of $\pi$ arises at $a_s = 0$. To remove this discontinuity and to include the full range $(-\pi, \pi]$ of $\tan^{-1}(b_s/a_s)$ in the simulation, we implement the following conditions in the codes:
\begin{enumerate}

\item[(i)] $0 \leq \Phi_s \leq \pi/2$ for $a_s \geq 0$, $b_s \geq 0$;
\item[(ii)] $\pi/2 < \Phi_s \leq \pi$ for $a_s < 0$, $b_s \geq 0$;
\item[(iii)] $-\pi < \Phi_s < -\pi/2$ for $a_s < 0$, $b_s < 0$;
\item[(iv)] $-\pi/2 \leq \Phi_s < 0$ for $a_s \geq 0$, $b_s < 0$;
\end{enumerate}
where, $\Phi_s = \tan^{-1}(b_s/a_s)$. It is to be noticed that the representation of Eq. (\ref{Amp_a+ib=rexp}) does not contradict with the previously described complex amplitude vector convention; because, not $\left( a_s^2 + b_s^2 \right)^{1/2}$, but the entire quantity $\left( a_s^2 + b_s^2 \right)^{1/2} e^{i \tan^{-1}(b_s/a_s)}$ is the $s$-component of the complex amplitude vector.

Nevertheless, our complex amplitude and phase conventions can readily put forward the wavefront-dislocation interpretation as well for appropriate cases; such as, for systems involving optical singularities \cite{Gbur}.

\section{Equations (\ref{H=kcrossE}--\ref{k.E=0}) with Complex Amplitude Vectors}\label{App_CompAmpConditions}

Here we show that the complex field amplitude vectors in the form of Eq. (\ref{Amp_Eform}) satisfy Eqs. (\ref{H=kcrossE}--\ref{k.E=0}). All quantities here are at the surface $S$ and are expressed in terms of the $S$ coordinate system. So we omit the subscript $S$ and the superscript `$(S)$' for simplicity.

For each $j = i,r,t$, we first consider two different electric fields with real amplitude vectors
\begin{eqnarray}
&& \boldsymbol{\mathcal{E}}_j^a = a_{jx} \, \hat{\mathbf{x}} + a_{jy} \, \hat{\mathbf{y}} + a_{jz} \, \hat{\mathbf{z}}; \label{CCond_Eja}\\
&& \boldsymbol{\mathcal{E}}_j^b = b_{jx} \, \hat{\mathbf{x}} + b_{jy} \, \hat{\mathbf{y}} + b_{jz} \, \hat{\mathbf{z}}; \label{CCond_Ejb}
\end{eqnarray}
having the same wavevector
\begin{equation}
\mathbf{k}_j = k_{jx} \, \hat{\mathbf{x}} + k_{jy} \, \hat{\mathbf{y}} + k_{jz} \, \hat{\mathbf{z}}.
\end{equation}
Both $\boldsymbol{\mathcal{E}}_j^a$ and $\boldsymbol{\mathcal{E}}_j^b$ are orthogonal to $\mathbf{k}_j$, thus satisfying
\begin{eqnarray}
\mathbf{k}_j \cdot \boldsymbol{\mathcal{E}}_j^a = 0;
\hspace{1em}
\mathbf{k}_j \cdot \boldsymbol{\mathcal{E}}_j^b = 0.
\end{eqnarray}
Multiplying $\mathbf{k}_j \cdot \boldsymbol{\mathcal{E}}_j^b$ by $i = e^{i \pi/2}$, and adding to $\mathbf{k}_j \cdot \boldsymbol{\mathcal{E}}_j^a$, we get
\begin{eqnarray}
&\mathbf{k}_j \cdot \boldsymbol{\mathcal{E}}_j = 0;
\hspace{1em}
\boldsymbol{\mathcal{E}}_j = \boldsymbol{\mathcal{E}}_j^a + i \, \boldsymbol{\mathcal{E}}_j^b = \boldsymbol{\mathcal{E}}_j^a + e^{i\frac{\pi}{2}} \, \boldsymbol{\mathcal{E}}_j^b \, . \label{CCond_k.E}
\end{eqnarray}
Thus, the single complex electric field amplitude vector $\boldsymbol{\mathcal{E}}_j$ is orthogonal to $\mathbf{k}_j$, verifying Eq. (\ref{k.E=0}). This orthogonality physically signifies that, in 3D space, the wavevector $\mathbf{k}_j$ is orthogonal to the plane of the polarization ellipse represented by the complex amplitude vector $\boldsymbol{\mathcal{E}}_j$.

We now consider the real magnetic field amplitude vectors $\boldsymbol{\mathcal{H}}_j^a$, $\boldsymbol{\mathcal{H}}_j^b$ corresponding to $\boldsymbol{\mathcal{E}}_j^a$, $\boldsymbol{\mathcal{E}}_j^b$; given by Eq. (\ref{H=kcrossE}) as
\begin{subequations}
\begin{eqnarray}
\boldsymbol{\mathcal{H}}_{j}^a = \left( \mathbf{k}_{j} \times \boldsymbol{\mathcal{E}}_{j}^a \right)/\omega\mu_0; \label{CCond_Hja} \\
\boldsymbol{\mathcal{H}}_{j}^b = \left( \mathbf{k}_{j} \times \boldsymbol{\mathcal{E}}_{j}^b \right)/\omega\mu_0. \label{CCond_Hjb}
\end{eqnarray}
\end{subequations}
Multiplying Eq. (\ref{CCond_Hjb}) by $i = e^{i \pi/2}$, and adding to Eq. (\ref{CCond_Hja}), we get
\begin{equation}
\boldsymbol{\mathcal{H}}_{j} = \left( \mathbf{k}_{j} \times \boldsymbol{\mathcal{E}}_{j} \right)/\omega\mu_0;
\hspace{1em}
\boldsymbol{\mathcal{H}}_j = \boldsymbol{\mathcal{H}}_j^a + i \, \boldsymbol{\mathcal{H}}_j^b = \boldsymbol{\mathcal{H}}_j^a + e^{i\frac{\pi}{2}} \, \boldsymbol{\mathcal{H}}_j^b. \label{CCond_H=kXE}
\end{equation}
Thus, the single complex magnetic field amplitude vector $\boldsymbol{\mathcal{H}}_j$, that corresponds to the single complex electric field amplitude vector $\boldsymbol{\mathcal{E}}_j$ [Eq. (\ref{CCond_k.E})], satisfies Eq. (\ref{H=kcrossE}).

The $s' = x, y$ components of $\boldsymbol{\mathcal{E}}_j^a$, $\boldsymbol{\mathcal{E}}_j^b$, for $j = i, r, t$, satisfy Eqs. (\ref{Exbc}, \ref{Eybc}):
\begin{subequations}
\begin{eqnarray}
a_{i s'} + a_{r s'} = a_{t s'}; &\hspace{1em} &\mbox{[Eqs. (\ref{Exbc}, \ref{Eybc}) for $\boldsymbol{\mathcal{E}}_j^a$ ]}; \label{CCond_asbc} \\
b_{i s'} + b_{r s'} = b_{t s'}; &\hspace{1em} &\mbox{[Eqs. (\ref{Exbc}, \ref{Eybc}) for $\boldsymbol{\mathcal{E}}_j^b$ ]}. \label{CCond_bsbc}
\end{eqnarray}
\end{subequations}
where, the component notations of Eqs. (\ref{CCond_Eja}, \ref{CCond_Ejb}) are used.
Multiplying Eq. (\ref{CCond_bsbc}) by $i = e^{i \pi/2}$, and adding to Eq. (\ref{CCond_asbc}), we get
\begin{eqnarray}
& \left(a_{i s'} + i\, b_{i s'}\right) + \left(a_{r s'} + i\, b_{r s'}\right) = \left(a_{t s'} + i\, b_{t s'}\right);& \nonumber \\
& \mbox{or,} \hspace{1em} \mathcal{E}_{i s'} + \mathcal{E}_{r s'} = \mathcal{E}_{t s'}; &
\end{eqnarray}
where, the $\mathcal{E}_{j s'} = a_{j s'} + i\, b_{j s'}$ terms are the $s'$-components of $\boldsymbol{\mathcal{E}}_j$ [Eq. (\ref{CCond_k.E})]. This verifies Eqs. (\ref{Exbc}, \ref{Eybc}) for the complex $\boldsymbol{\mathcal{E}}_j$. In a similar way, by considering component representations of $\boldsymbol{\mathcal{H}}_j^a$, $\boldsymbol{\mathcal{H}}_j^b$, $\boldsymbol{\mathcal{H}}_j$, Eqs. (\ref{Hxbc}, \ref{Hybc}) are verified for the complex $\boldsymbol{\mathcal{H}}_j$ [Eq. (\ref{CCond_H=kXE})].

\section{Derivation of Eqs. (\ref{EjSS}--\ref{tMat})}\label{App_CentralResult}

Here we show the derivation of Eqs. (\ref{EjSS}--\ref{tMat}). Here, all amplitude vector components are complex in general, as discussed in Appendices \ref{App_Amplitude}, \ref{App_CompAmpConditions}. 
Also, like Appendix \ref{App_CompAmpConditions}, we omit the subscript $S$ and the superscript `$(S)$' for simplicity.

Expanding Eq. (\ref{H=kcrossE}) for $j=i,r,t$ using Eq. (\ref{kjSS}), and then using in Eqs. (\ref{Hxbc}, \ref{Hybc}), we get
\begin{eqnarray}
(k_{iy}\mathcal{E}_{iz} - k_{iz}\mathcal{E}_{iy}) + (k_{iy}\mathcal{E}_{rz} + k_{iz}\mathcal{E}_{ry}) \hspace{5em}&&\nonumber\\ 
= (k_{iy}\mathcal{E}_{tz} - k_{tz}\mathcal{E}_{ty});\hspace{1em}&&
\label{A_BC_Hx}\\
(k_{iz}\mathcal{E}_{ix} - k_{ix}\mathcal{E}_{iz}) - (k_{iz}\mathcal{E}_{rx} + k_{ix}\mathcal{E}_{rz}) \hspace{5em}&&\nonumber\\ 
= (k_{tz}\mathcal{E}_{tx} - k_{ix}\mathcal{E}_{tz}).\hspace{1em}&&
\label{A_BC_Hy}
\end{eqnarray}
Expanding Eq. (\ref{k.E=0}) for $j=i,r,t$ using Eq. (\ref{kjSS}), and rearranging, we get
\begin{eqnarray}
\mathcal{E}_{iz} &=& -(k_{ix}\mathcal{E}_{ix} + k_{iy}\mathcal{E}_{iy})/k_{iz};\label{A_E1z}\\
\mathcal{E}_{rz} &=& (k_{ix}\mathcal{E}_{rx} + k_{iy}\mathcal{E}_{ry})/k_{iz};\label{A_Erz}\\
\mathcal{E}_{tz} &=& -(k_{ix}\mathcal{E}_{tx} + k_{iy}\mathcal{E}_{ty})/k_{tz}.\label{A_Etz}
\end{eqnarray}
Using Eqs. (\ref{A_E1z}--\ref{A_Etz}) in Eqs. (\ref{A_BC_Hx}, \ref{A_BC_Hy}), we get
\begin{eqnarray}
k_{ix} k_{iy} (\mathcal{E}_{ix} - \mathcal{E}_{rx})/k_{iz} + \left( k_{iz} + k_{iy}^2/k_{iz} \right)(\mathcal{E}_{iy} - \mathcal{E}_{ry}) \hspace{2em}&&\nonumber\\
 = k_{ix} k_{iy} \,\mathcal{E}_{tx}/k_{tz} + \left( k_{tz} + k_{iy}^2/k_{tz} \right)\mathcal{E}_{ty};\hspace{2em}&&\label{A_B9b}\\
\left( k_{iz} + k_{ix}^2/k_{iz} \right)(\mathcal{E}_{ix} - \mathcal{E}_{rx}) + k_{ix} k_{iy} (\mathcal{E}_{iy} - \mathcal{E}_{ry})/k_{iz} \hspace{2em}&& \nonumber\\
 = \left( k_{tz} + k_{ix}^2 / k_{tz} \right)\mathcal{E}_{tx} + k_{ix} k_{iy}\,\mathcal{E}_{ty}/k_{tz}. \hspace{2em}&&\label{A_B9a}
\end{eqnarray}
Substituting $\mathcal{E}_{tx}$ and $\mathcal{E}_{ty}$ from Eqs. (\ref{Exbc}, \ref{Eybc}) to Eqs. (\ref{A_B9b}, \ref{A_B9a}), and rearranging, we get
\begin{equation}
\begin{bmatrix}
C_{xy} & C_{yy} \\
C_{xx} & C_{xy} \end{bmatrix}
\begin{bmatrix}
\mathcal{E}_{rx} \\
\mathcal{E}_{ry} \end{bmatrix}
= 
\begin{bmatrix}
D_{xy} & D_{yy} \\
D_{xx} & D_{xy} \end{bmatrix}
\begin{bmatrix}
\mathcal{E}_{ix} \\
\mathcal{E}_{iy} \end{bmatrix};
\label{A_CD_matEq}
\end{equation}
\begin{subequations}
\begin{eqnarray}
\mbox{where,} \hspace{1em} && C_{xx} = (k_{tz} + k_{iz}) \left(k_{ix}^2 /k_{tz}k_{iz} + 1 \right);\hspace{1em}\\
&& C_{yy} = (k_{tz} + k_{iz}) \left( k_{iy}^2 /k_{tz}k_{iz} + 1 \right);\\
&& C_{xy} = k_{ix} k_{iy} (k_{tz} + k_{iz})/k_{tz}k_{iz};\\
&& D_{xx} = (k_{tz} - k_{iz}) \left( k_{ix}^2 /k_{tz}k_{iz} - 1\right);\\
&& D_{yy} = (k_{tz} - k_{iz}) \left( k_{iy}^2 /k_{tz}k_{iz} - 1 \right);\\
&& D_{xy} = k_{ix} k_{iy} (k_{tz} - k_{iz})/k_{tz}k_{iz}.
\end{eqnarray}
\end{subequations}
\if{false}
\begin{eqnarray}
\mbox{where,}&& C_{xx} = (k_{iz} + k_{tz}) + k_{ix}^2 (k_{tz} + k_{iz})/k_{tz}k_{iz};\nonumber\\
&& C_{yy} = (k_{iz} + k_{tz}) + k_{iy}^2 (k_{tz} + k_{iz})/k_{tz}k_{iz};\nonumber\\
&& C_{xy} = k_{ix} k_{iy} (k_{tz} + k_{iz})/k_{tz}k_{iz};\nonumber\\
&& D_{xx} = (k_{iz} - k_{tz}) + k_{ix}^2 (k_{tz} - k_{iz})/k_{tz}k_{iz};\nonumber\\
&& D_{yy} = (k_{iz} - k_{tz}) + k_{iy}^2 (k_{tz} - k_{iz})/k_{tz}k_{iz};\nonumber\\
&& D_{xy} = k_{ix} k_{iy} (k_{tz} - k_{iz})/k_{tz}k_{iz}.\nonumber
\end{eqnarray}
\fi
We have verified that $\begin{bmatrix}
C_{xy} & C_{yy} \\
C_{xx} & C_{xy} \end{bmatrix}$ is non-singular. So, we solve Eq. (\ref{A_CD_matEq}) for $\begin{bmatrix}
\mathcal{E}_{rx} \\
\mathcal{E}_{ry} \end{bmatrix}$ and obtain
\begin{eqnarray}
\begin{bmatrix}
\mathcal{E}_{rx} \\
\mathcal{E}_{ry} \end{bmatrix}
&=& 
\begin{bmatrix}
C_{xy} & C_{yy} \\
C_{xx} & C_{xy} \end{bmatrix}
^{-1}
\begin{bmatrix}
D_{xy} & D_{yy} \\
D_{xx} & D_{xy} \end{bmatrix}
\begin{bmatrix}
\mathcal{E}_{ix} \\
\mathcal{E}_{iy} \end{bmatrix}
\nonumber\\
&=& A_0 \begin{bmatrix}
A_{11} & A_{xy} \\
A_{xy} & -A_{10} \end{bmatrix}
\begin{bmatrix}
\mathcal{E}_{ix} \\
\mathcal{E}_{iy} \end{bmatrix}
;\label{A_ErxEryMatA}
\end{eqnarray}
where, $A_0$, $A_{11}$, $A_{xy}$ and $A_{10}$ are given by Eqs. (\ref{Aterms}).
Then, using Eqs. (\ref{Exbc}, \ref{Eybc}, \ref{A_ErxEryMatA}), $\mathcal{E}_{tx}$ and $\mathcal{E}_{ty}$ can be expressed as
\begin{equation}
\begin{bmatrix}
\mathcal{E}_{tx} \\
\mathcal{E}_{ty} \end{bmatrix}
 = 
\begin{bmatrix}
1 + A_0 A_{11}  &  A_0 A_{xy} \\
A_0 A_{xy}  &  1 - A_0 A_{10} \end{bmatrix}
\begin{bmatrix}
\mathcal{E}_{ix} \\
\mathcal{E}_{iy} \end{bmatrix}
.\label{A_EtxEtyMatA}
\end{equation}
Finally, using Eqs. (\ref{A_ErxEryMatA}, \ref{A_EtxEtyMatA}) in Eqs. (\ref{A_Erz}, \ref{A_Etz}), and simplifying by using Eqs. (\ref{Aterms}, \ref{A_E1z}), $\mathcal{E}_{rz}$ and $\mathcal{E}_{tz}$ are obtained as
\begin{eqnarray}
\mathcal{E}_{rz} &=& -A_0 A_{01}\mathcal{E}_{iz};\label{A_ErzA}\\
\mathcal{E}_{tz} &=& \dfrac{k_{iz}}{k_{tz}} (1 + A_0 A_{01})\,\mathcal{E}_{iz};\label{A_EtzA}
\end{eqnarray}
where, $A_{01}$ is given by Eq. (\ref{Apq}). Then, Eqs. (\ref{A_ErxEryMatA}, \ref{A_ErzA}) are combined and written as $\boldsymbol{\mathcal{E}}_r = \tilde{\mathbf{r}}\, \boldsymbol{\mathcal{E}}_i$ (Eq. (\ref{EjSS}) for $j' = r$), where $\tilde{\mathbf{r}}$ is the reflection coefficient matrix [Eq. (\ref{rMat})]. Similarly, Eqs. (\ref{A_EtxEtyMatA}, \ref{A_EtzA}) are combined and written as $\boldsymbol{\mathcal{E}}_t = \tilde{\mathbf{t}}\, \boldsymbol{\mathcal{E}}_i$ (Eq. (\ref{EjSS}) for $j' = t$), where $\tilde{\mathbf{t}}$ is the transmission coefficient matrix [Eq. (\ref{tMat})].


\section{Derivation of Eqs. (\ref{EiII_model}--\ref{RIPPgI})} \label{App_Ei_Model}

The transformation of the wavevector $\mathbf{k}_{0}^{(I)}$ [Eq. (\ref{k0I=n1kz})] to the wavevector $\tilde{\mathbf{k}}_{iI}^{(I)}$ [Eq. (\ref{kiII=n1kr})] can be understood in terms of geometrical rotations. We consider a coordinate system $I'$ such that $\hat{\mathbf{z}}^{(I')} = \hat{\mathbf{z}}^{(I)}$ and $\hat{\mathbf{x}}^{(I')} = \cos\phi_I \, \hat{\mathbf{x}}^{(I)} + \sin\phi_I \, \hat{\mathbf{y}}^{(I)}$. The transformation of any vector from the $I$ coordinate system to the $I'$ coordinate system is obtained by the application of the rotation matrix $\tilde{\mathbf{R}}_{I'I}$ [Eq. (\ref{RIIP})]. Then, applying a rotation $\tilde{\mathbf{R}}_{I'I''}$ [Eq. (\ref{RIPPgI})] to the transformed wavevector $\mathbf{k}_{0}^{(I')} = \tilde{\mathbf{R}}_{I'I}\, \mathbf{k}_{0}^{(I)}$, and then transforming it back to the $I$ coordinate system, we obtain the wavevector $\tilde{\mathbf{k}}_{iI}^{(I)}$. Thus,
\begin{equation}
\tilde{\mathbf{k}}_{iI}^{(I)} = \tilde{\mathbf{R}}_{II'} \tilde{\mathbf{R}}_{I'I''} \tilde{\mathbf{R}}_{I'I} \, \mathbf{k}_{0}^{(I)}.
\end{equation}

\begin{figure}[t]
\centering
\includegraphics[width = 0.65\linewidth]{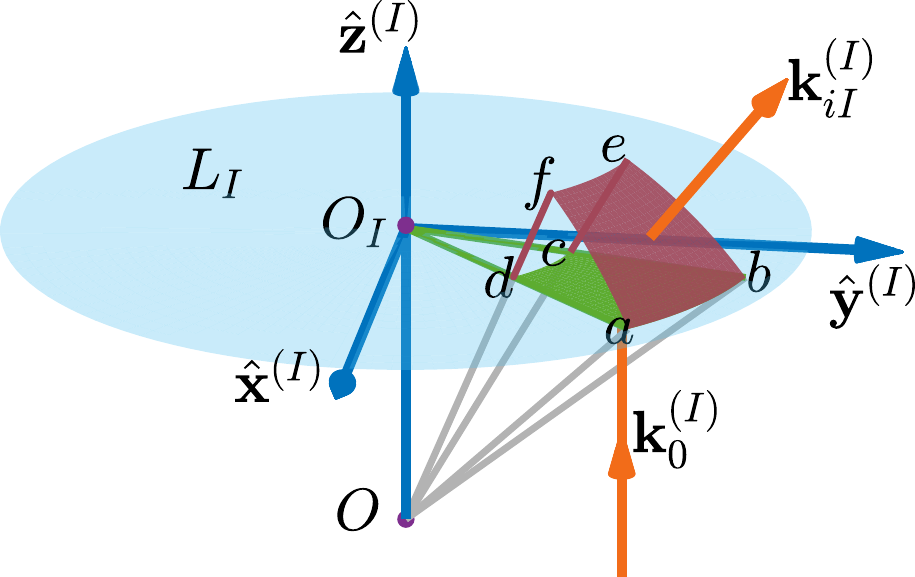}
\caption{\small{3D model on the operation of the lens $L_I$ of Fig. \ref{fig_setup}. The initial wavevector $\mathbf{k}_0^{(I)}$ [Eq. (\ref{k0I=n1kz})] is transformed to the wavevector $\mathbf{k}_{iI}^{(I)}$ [Eqs. (\ref{kiII=n1kr})] by the operation of the lens $L_I$. Correspondingly, the plane-polar surface element $abcd$ (area $ = dS_0$) of the initial collimated wave just before $L_I$ is transformed to a spherical surface element $abef$ (area $ = dS_I$) of the diverging wave just after $L_I$. Here, 
$O_I d = O_I c = \rho^{(I)}$; 
$da = cb = d\rho^{(I)}$; 
$\angle d O_I c = \angle a O_I b = d\phi_I$; 
$ Oa = Ob = Oe = Of = r_I $;
$\angle O_I O d = \angle O_I O c = \theta_I$; 
$\angle d O a = \angle c O b = d\theta_I$. The change in the electric field magnitude is given by the factor $g_I = \sqrt{dS_0/dS_I} = 1/\sqrt{\cos\theta_I}$ [Eq. (\ref{RIPPgI})].
}}
{\color{grey}{\rule{\linewidth}{1pt}}}
\label{fig_Lens}
\end{figure}

Now, no rotation of the local electric field amplitude vector about the local wavevector should happen in this transformation \cite{BARev}. This means that the same transformation $\tilde{\mathbf{R}}_{II'} \tilde{\mathbf{R}}_{I'I''} \tilde{\mathbf{R}}_{I'I}$ must be applied to the electric field amplitude vector $\boldsymbol{\mathcal{E}}_0^{(I)}$ [Eq. (\ref{E0I})] as well. However, there is also an overall change in the field amplitude value due to the change in the wavefront shape. We consider a plane-polar surface element of area $dS_0 = \rho^{(I)} d\rho^{(I)} d\phi_I$ on the planar wavefront just before the lens $L_I$ [Fig. \ref{fig_Lens}]. This element transforms to a spherical surface element of area $dS_I = r_I^2 \sin\theta_I d\theta_I d\phi_I$ after passing through $L_I$. We know that $r_I\sin\theta_I = \rho^{(I)}$ [Eq. (\ref{cs_thetaI})]; and we have verified geometrically that
\begin{equation}
r_I \, d\theta_I = \cos\theta_I \, d\rho^{(I)}
\end{equation}
(usual differentiation of $\rho^{(I)} = r_I\sin\theta_I$ is not applicable in this special scenario). Using these relations, we find that the area of the element changes by a factor $dS_I/dS_0 = \cos\theta_I$. So, the intensity changes by the inverse factor $1/\cos\theta_I$; and hence the electric field magnitude changes by the factor $g_I = 1/\sqrt{\cos\theta_I}$ [Eq. (\ref{RIPPgI})]. Thus, by applying the transformation $\tilde{\mathbf{R}}_{II'} \tilde{\mathbf{R}}_{I'I''} \tilde{\mathbf{R}}_{I'I}$ along with the amplitude modifying factor $g_I$ to $\boldsymbol{\mathcal{E}}_0^{(I)}$, we obtain the electric field amplitude vector $\tilde{\boldsymbol{ \mathcal{E}}}_{iI}^{(I)}$ [Eq. (\ref{EiII_model})].

We have verified the correctness of Eq. (\ref{EiII_model}) by utilizing energy conservation --- we have verified by using Eqs. (\ref{E0I}--\ref{E0yI}, \ref{EiII_model}--\ref{RIPPgI}) that the total powers of the fields $\mathbf{E}_0^{(I)}$ (before $L_I$) and $\tilde{\mathbf{E}}_{iI}^{(I)}$ (just after $L_I$) are equal [Eq. (\ref{P0=PiI})].

\vspace{1em}

\section{A Note on the Factor \textit{g}$_{TS}$ of Eq. (\ref{EtTT})} \label{App_gTS}

The amplitude modifying multiplicative factor $g_{TS}$ is unity for the Fourier decomposition case [Subsection \ref{Sub_FourierRecombination}]. However, for the wavefront-surface-element decomposition case it has a complicated form.

We consider the spherically diverging incident wave of our simulated model as an example. Though the reflected wave in this case retains the spherically diverging geometry, the transmitted wave does not. We have verified that, each surface element on any given transmitted wavefront has a unique set of polar radius of curvature, polar center of curvature, azimuthal radius of curvature and azimuthal center of curvature The collection of the polar centers of curvature of all surface elements form a 3D caustic surface, from which the transmitted wave appears to be emitted.

Now, for one such surface element under consideration, we denote the polar and azimuthal radii of curvature at the surface $S$ as $r_{\theta S}$ and $r_{\phi S}$ respectively; and those at the surface $T$ as $r_{\theta T}$ and $r_{\phi T}$ respectively. Then, the change in area of the element due to the propagation is given by the factor $r_{\theta T} r_{\phi T}/r_{\theta S} r_{\phi S}$. The change in intensity is then given by the inverse factor $r_{\theta S} r_{\phi S}/r_{\theta T} r_{\phi T}$; and hence we obtain the amplitude modifying multiplicative factor as $g_{TS} = (r_{\theta S} r_{\phi S}/r_{\theta T} r_{\phi T})^{1/2}$.


\vspace{1em}

\section{Geometric Phase Consideration} \label{App_GeoPhase}

In this section, we have conveniently omitted the phase terms of the form $\left( \mathbf{k} \cdot \Delta\mathbf{r} - \omega t + \Phi_O \right)$ for the simplicity of the discussions. Here, we demonstrate the nature of the geometric phase involved in our formalism by taking the example of a $\hat{\boldsymbol{\sigma}}^+$ spin-polarized $\boldsymbol{\mathcal{E}}_0^{(I)}$ field [Eq. (\ref{E0I})]. We take $\theta_E = 45^\circ$, $\Phi_E = \pi/2$ in Eqs. (\ref{E0I}--\ref{E0yI}), and obtain 
\begin{eqnarray}
\boldsymbol{\mathcal{E}}_0^{(I)} &=& \mathcal{E}_{00} \, e^{-\rho^{(I) \, 2}/w_0^2} \; \frac{1}{\sqrt{2}} \left( \hat{\mathbf{x}}^{(I)} + i\, \hat{\mathbf{y}}^{(I)} \right)\nonumber\\
&\equiv & \mathcal{E}_{00} \, e^{-\rho^{(I) \, 2}/w_0^2} \; \frac{1}{\sqrt{2}} \begin{bmatrix}
1 \\ i \\ 0
\end{bmatrix}. \label{E0I_modelG}
\end{eqnarray}
After passing through the lens $L_I$, the functional form of the wavevector $\tilde{\mathbf{k}}_{iI}^{(I)}$ is obtained by Eqs. (\ref{kiII=n1kr}, \ref{rII}); and, via Eq. (\ref{EiII_model}), the field $\boldsymbol{\mathcal{E}}_0^{(I)}$ transforms to
\begin{eqnarray}
\tilde{\boldsymbol{\mathcal{E}}}_{iI}^{(I)} &=& \mathcal{E}_{00} \, e^{-\rho^{(I) \, 2}/w_0^2} \; \frac{1}{\sqrt{2 \cos\theta_I}} \begin{bmatrix}
\tilde{e}_{ixI}^{(I)}
\\ \tilde{e}_{iyI}^{(I)} \\ \tilde{e}_{izI}^{(I)}
\end{bmatrix};\\
\hspace{-1em} \mbox{where,} && \tilde{e}_{ixI}^{(I)} = \left( \cos\theta_I \cos^2\phi_I + \sin^2\phi_I \right) \nonumber\\ 
&& \hspace{5em} - \, i \cos\phi_I \sin\phi_I (1 - \cos\theta_I);\\
&& \tilde{e}_{iyI}^{(I)} = -\cos\phi_I \sin\phi_I (1 - \cos\theta_I) \nonumber\\ 
&& \hspace{5em} + \, i \left( \cos\theta_I \sin^2\phi_I + \cos^2\phi_I \right);\\
&& \tilde{e}_{izI}^{(I)} = -\sin\theta_I \, e^{i\phi_I}.
\end{eqnarray}

We now consider two wavefront-surface-elements at the surface $I$. 
We identify each element as element-$p$ ($p = 1, 2$); located at $\left( x_p^{(I)}, y_p^{(I)}, 0^{(I)} \right)$; with the corresponding quantities $\rho_p^{(I)}$, $r_{Ip}$, $\theta_{Ip}$, $\phi_{Ip}$, $\hat{\mathbf{r}}_{Ip}^{(I)}$ determined by Eqs. (\ref{rhoI}--\ref{cs_thetaI}, \ref{rII}). The constituent plane wavevector and electric field amplitude vector corresponding to each of these surface-element plane-waves are given by the local values of $\tilde{\mathbf{k}}_{iI}^{(I)}$ and $\tilde{\boldsymbol{\mathcal{E}}}_{iI}^{(I)}$ as
\begin{eqnarray}
& \mathbf{k}_{iIp}^{(I)} = n_1 k \, \hat{\mathbf{r}}_{Ip}^{(I)}; &\\
& \hspace{-5em} \boldsymbol{\mathcal{E}}_{iIp}^{(I)} = \mathcal{E}_{00} \, e^{-\rho_p^{(I) \, 2}/w_0^2} \; \dfrac{1}{\sqrt{2 \cos\theta_{Ip}}} & \nonumber \\ 
& \hspace{4em} \times \left( e_{ixIp}^{(I)} \, \hat{\mathbf{x}}^{(I)} + e_{iyIp}^{(I)} \, \hat{\mathbf{y}}^{(I)} + e_{izIp}^{(I)} \, \hat{\mathbf{z}}^{(I)} \right); & 
\end{eqnarray}
where, $e_{isIp}^{(I)}$ ($s = x, y, z$) are the local values of $\tilde{e}_{isI}^{(I)}$ for $\theta_{Ip}, \phi_{Ip}$.

As per the transformation of Eq. (\ref{EiII_model}), each field $\boldsymbol{\mathcal{E}}_{iIp}^{(I)}$ retains the initial $\hat{\boldsymbol{\sigma}}^+$ polarization [Eq. (\ref{E0I_modelG})] at the plane of the element-$p$ --- the plane normal to  $\mathbf{k}_{iIp}^{(I)}$. However, each $\boldsymbol{\mathcal{E}}_{iIp}^{(I)}$ creates a different elliptical polarization projection on the surface $I$, given by
\begin{equation}
\boldsymbol{\mathcal{E}}_{iIp}^{\prime (I)} = \dfrac{\mathcal{E}_{00} \, e^{-\rho_p^{(I) \, 2}/w_0^2}}{\sqrt{2 \cos\theta_{Ip}}} 
\left( e_{ixIp}^{(I)} \, \hat{\mathbf{x}}^{(I)} + e_{iyIp}^{(I)} \, \hat{\mathbf{y}}^{(I)} \right).
\label{EiIppI_G}
\end{equation}
Equation (\ref{EiIppI_G}) represents the projected polarization ellipses of Fig. \ref{fig_field_profiles}(c) for all the wavefront-surface-elements at the surface $I$. 
The different projected polarization ellipses correspond to different states on the Poincar\'e sphere; 
and additionally, these different projected fields differ by geometric phase factors which can be calculated explicitly.

\begin{figure}[t]
\centering
\includegraphics[width = 0.72\linewidth]{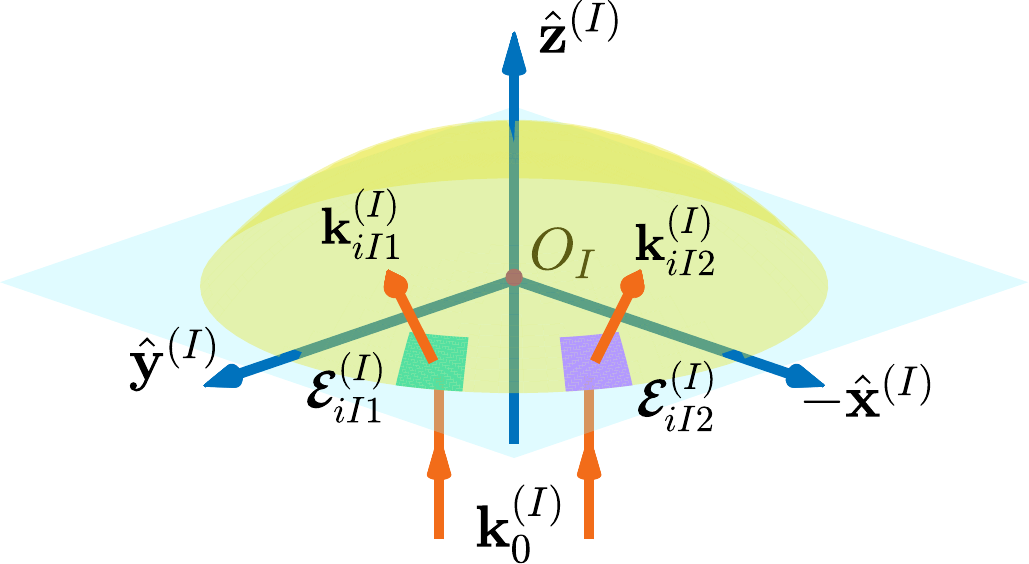}
\caption{\small{
Demonstration of geometric phase calculation [Appendix \ref{App_GeoPhase}] by using two surface-element fields $\boldsymbol{\mathcal{E}}_{iIp}^{(I)}$ ($p = 1,2$) on the same wavefront.
}}
{\color{grey}{\rule{\linewidth}{1pt}}}
\label{fig_TwoSurfElem}
\end{figure}

For the purpose of this demonstration with the previously-mentioned two elements ($p = 1,2$), 
it is now sufficient to proceed further by considering some particularly convenient values of $\theta_{Ip}$ and $\phi_{Ip}$ to evade the involvement of the general conventions discussed in Appendix \ref{App_Amplitude}.
We consider the values $\theta_{I1} = \theta_{I2} = 60^\circ$, $\phi_{I1} = 120^\circ$ and $\phi_{I2} = 150^\circ$ (the coordinates $\left( x_p^{(I)}, y_p^{(I)} \right)$ are adjusted accordingly, based on Eqs. (\ref{rhoI}--\ref{cs_thetaI}) ). Thus, both the elements are on the same wavefront [Fig. \ref{fig_TwoSurfElem}] \cite{FOOTNOTE_2elem}, whose circular intercept at the surface $I$ has a radius $\rho_{1}^{(I)} = \rho_2^{(I)} = \sqrt{3} \, |f|$ (using Eq. (\ref{cs_thetaI}) ). Then, by using Eq. (\ref{EiIppI_G}), the two projected fields $\boldsymbol{\mathcal{E}}_{iIp}^{\prime (I)}$ are obtained as
\begin{eqnarray}
&&\hspace{-2em}\boldsymbol{\mathcal{E}}_{iI1}^{\prime (I)} = \mathcal{E}_0 \left[ \left(7 + i \sqrt{3}\right) \hat{\mathbf{x}}^{(I)} + \left(\sqrt{3} + i\, 5\right) \hat{\mathbf{y}}^{(I)} \right]; \label{EiI1pI_G} \\
&&\hspace{-2em}\boldsymbol{\mathcal{E}}_{iI2}^{\prime (I)} = \mathcal{E}_0 \left[ \left(5 + i \sqrt{3}\right) \hat{\mathbf{x}}^{(I)} + \left(\sqrt{3} + i\, 7\right) \hat{\mathbf{y}}^{(I)} \right]; \label{EiI2pI_G} \\
&& \mbox{where,} \hspace{1em} \mathcal{E}_0 = (\mathcal{E}_{00}/8)\, e^{-3 f^2/w_0^2}.\label{E0def_G}
\end{eqnarray}

We now consider a virtual superposition of these two fields \cite{FOOTNOTE_Superposition} in the form 
\begin{equation}
\boldsymbol{\mathcal{E}}_{iI}^{\prime (I)} = \boldsymbol{\mathcal{E}}_{iI1}^{\prime (I)} + e^{-i \delta} \boldsymbol{\mathcal{E}}_{iI2}^{\prime (I)}. \label{EiIpI_G}
\end{equation}
According to Pancharatnam's criterion \cite{P1956}, there exists a phase value $\delta = \delta_G$, for which the intensity of the superposed field $\boldsymbol{\mathcal{E}}_{iI}^{\prime (I)}$ acquires the maximum value. The field $e^{-i \delta_G} \boldsymbol{\mathcal{E}}_{iI2}^{\prime (I)}$ is then identified to be `in phase' 
with $\boldsymbol{\mathcal{E}}_{iI1}^{\prime (I)}$; and consequently, the field $\boldsymbol{\mathcal{E}}_{iI2}^{\prime (I)}$ is identified to lead the field $\boldsymbol{\mathcal{E}}_{iI1}^{\prime (I)}$ by a geometric phase $\delta_G$.

Using Eqs. (\ref{EiI1pI_G}, \ref{EiI2pI_G}) in Eq. (\ref{EiIpI_G}), we get
\begin{equation}
\left| \boldsymbol{\mathcal{E}}_{iI}^{\prime (I)} \right|^2 = 8\, \mathcal{E}_0^2 \left( 20 + 19 \cos\delta + \sqrt{3} \sin\delta \right).
\end{equation}
Then, using the conditions 
\begin{equation}
\left. d\, | \boldsymbol{\mathcal{E}}_{iI}^{\prime (I)} |^2 / d  \delta \right|_{\delta = \delta_G} = 0; 
\hspace{1em} 
\left. d^2\, | \boldsymbol{\mathcal{E}}_{iI}^{\prime (I)} |^2 / d  \delta^2 \right|_{\delta = \delta_G} < 0 ;
\end{equation}
we obtain the geometric phase as $\delta_G = \tan^{-1}\left(\sqrt{3}/19\right)$. In this way, the geometric phase difference between the two projected constituent plane-wave fields are calculated.

The above demonstration is based on the field $\tilde{\boldsymbol{\mathcal{E}}}_{iI}^{(I)}$. However, since the exact 3D expressions of all the fields $\tilde{\boldsymbol{\mathcal{E}}}_{jC}^{(C)}$ are available in our formalism, the above method can be straightforwardly utilized to find the geometric phase characteristics of all the other fields at the various surfaces.

In comparison, for the Fourier decomposition case, the constituent plane-wave fields exist everywhere at any given $C$ surface [Fig. \ref{fig_FourierComps}]. They posses surface-element characteristics on the surface of the unit-sphere of directions $\left\{ \hat{\mathbf{k}}_{jC}^{(C)} \right\}$, $\hat{\mathbf{k}}_{jC}^{(C)} = \tilde{\mathbf{k}}_{jC}^{(C)}/n k$, in the $\tilde{\mathbf{k}}_{jC}^{(C)}$ momentum space (as understood from Figure 2 of Ref. \cite{BARev}).  
Nevertheless, the same method as above is utilized to determine the geometric phase characteristics in the Fourier decomposition case as well. 

The above discussion shows that the geometric phase information is an additional feature of our formalism --- already contained in the exact 3D electric field expressions. One can choose to explicitly calculate these geometric phase characteristics; but this calculation is never a necessity for the application of our formalism.



\vspace{2em}

\begin{acknowledgments}
A.D. thanks Council of Scientific and Industrial Research, India for research fellowship (CSIR-JRF); N.K.V. thanks Science and Engineering Research Board, Department of Science and Technology (SERB, DST), India for financial support.
\end{acknowledgments}


\begin{thebibliography}{99}

\bibitem{Jackson}
J. D. Jackson, \textit{Classical Electrodynamics}, 3rd ed. (John Wiley \& Sons (Asia) Pte. Ltd, Singapore, 1999).

\bibitem{SalehTeich}
B. E. A. Saleh and M. C. Teich, \textit{Fundamentals of Photonics}, 2nd ed. (John Wiley \& Sons, Inc., NJ, 2007).

\bibitem{BornWolf}
M. Born and E. Wolf, \textit{Principles of Optics}, 7th ed. (Cambridge University Press, Cambridge, 1999).

\bibitem{Gbur}
G. J. Gbur, \textit{Singular Optics} (CRC Press, Taylor \& Francis Group, LLC, FL, 2017).



\bibitem{Poynting}
J. H. Poynting, The wave motion of a revolving shaft, and a suggestion as to the angular momentum in a beam of circularly polarised light, Proc. R. Soc. A \textbf{82}, 560 (1909).

\bibitem{RABeth}
R. A. Beth, Mechanical detection and measurement of the angular momentum of light, Phys. Rev. \textbf{50}, 115 (1936).

\bibitem{LAllen}
L. Allen, M. W. Beijersbergen, R. J. C. Spreeuw, and J. P. Woerdman, Orbital angular momentum of light and the transformation of Laguerre-Gaussian laser modes, Phys. Rev. A \textbf{45}, 8185 (1992).

\bibitem{HHe}
H. He, M. E. J. Friese, N. R. Heckenberg, and H. Rubinsztein-Dunlop, Direct observation of transfer of angular momentum to absorptive particles from a laser beam with a phase singularity, Phys. Rev. Lett. \textbf{75}, 826 (1995).



\bibitem{PA2000}
M. Padgett and A. Allen, Light with a twist in its tail, Contemporary Physics \textbf{41}, 275 (2000).

\bibitem{Yao}
A. M. Yao and M. J. Padgett, Orbital angular momentum: origins, behavior and applications, Adv. Opt. Photon. \textbf{3}, 161 (2011).

\bibitem{DennisVortex}
M. R. Dennis, K. O'Holleran, and M. J. Padgett, Progress in optics (Elsevier, Amsterdam, 2009) Chap. Singular Optics: Optical Vortices and Polarization Singularities, pp. 293--363.

\bibitem{BNRev}
K. Y. Bliokh and F. Nori, Transverse and longitudinal angular momenta of light, Physics Reports \textbf{592}, 1 (2015).



\bibitem{UriLevyRev}
U. Levy, Y. Silberberg, and N. Davidson, Mathematics of vectorial Gaussian beams, Adv. Opt. Photon. \textbf{11}, 828 (2019).

\bibitem{GH}
V. F. Goos and H. H\"anchen, Ein neuer und fundamentaler versuch zur totalreflexion, Ann. Physik \textbf{436}, 333 (1947).

\bibitem{Artmann}
K. Artmann, Berechnung der seitenversetzung des totalreflektierten strahles, Ann. Physik \textbf{437}, 87 (1948).

\bibitem{RaJW}
J. W. Ra, H. L. Bertoni, and L. B. Felsen, Reflection and transmission of beams at a dielectric interface, SIAM J. Appl. Math. \textbf{24}, 396 (1973).



\bibitem{AntarYM}
Y. M. Antar and W. M. Boerner, Gaussian beam interaction with a planar dielectric interface, Can. J. Phys. \textbf{52}, 962 (1974).

\bibitem{McGuirk}
M. McGuirk and C. K. Carniglia, An angular spectrum representation approach to the Goos-H\"anchen shift, J. Opt. Soc. Am. \textbf{67}, 103 (1977).

\bibitem{ChanCC}
C. C. Chan and C. Tamir, Angular shift of a Gaussian beam reflected near the Brewster angle, Opt. Lett. \textbf{10}, 378 (1985).

\bibitem{Porras}
M. A. Porras, Moment-method evaluation of the angular and lateral shifts of reflected light beams, Optics Communications \textbf{131}, 13 (1996).



\bibitem{AielloArXiv}
A. Aiello and J. P. Woerdman, Theory of angular Goos-H\"anchen shift near Brewster incidence, arXiv:0903.3730v2 [physics.optics] (2009).

\bibitem{Fedorov}
F. I. Fedorov, To the theory of total reflection, Dokl. Akad. Nauk SSSR \textbf{105}, 465 (1955), English translation available at http://master.basnet .by/congress2011/symposium/spbi.pdf.

\bibitem{Schilling}
H. Schilling, Die strahlversetzung bei der reflexion linear oder elliptisch polarisierter ebener wellen an der trennebene zwischen absorbierenden medien, Ann. Physik \textbf{471}, 122 (1965).

\bibitem{Imbert}
C. Imbert, Calculation and experimental proof of the transverse shift induced by total internal reflection of a circularly polarized light beam, Phys. Rev. D \textbf{5}, 787 (1972).



\bibitem{Player}
M. A. Player, Angular momentum balance and transverse shifts on reflection of light, J. Phys. A: Math. Gen. \textbf{20}, 3667 (1987).

\bibitem{FVG}
V. G. Fedoseyev, Conservation laws and transverse motion of energy on reflection and transmission of electromagnetic waves, J. Phys. A: Math. Gen. \textbf{21}, 2045 (1988).

\bibitem{Liberman}
V. S. Liberman and B. Y. Zel'dovich, Spin-orbit interaction of a photon in an inhomogeneous medium, Phys. Rev. A \textbf{46}, 5199 (1992).

\bibitem{Onoda}
M. Onoda, S. Murakami, and N. Nagaosa, Hall effect of light, Phys. Rev. Lett. \textbf{93}, 083901 (2004).



\bibitem{Bliokh2006}
K. Y. Bliokh and Y. P. Bliokh, Conservation of angular momentum, transverse shift, and spin Hall effect in reflection and refraction of an electromagnetic wave packet, Phys. Rev. Lett. \textbf{96}, 073903 (2006).

\bibitem{Bliokh2007}
K. Y. Bliokh and Y. P. Bliokh, Polarization, transverse shifts, and angular momentum conservation laws in partial reflection and refraction of an electromagnetic wave packet, Phys. Rev. E \textbf{75}, 066609 (2007).

\bibitem{HostenKwiat}
O. Hosten and P. Kwiat, Observation of the spin Hall effect of light via weak measurements, Science \textbf{319}, 787 (2008).

\bibitem{AielloArXiv2}
A. Aiello and J. P. Woerdman, The reflection of a Maxwell-Gaussian beam by a planar surface, arXiv:0710.1643v2 [physics.optics] (2007).



\bibitem{Aiello2008}
A. Aiello and J. P. Woerdman, Role of beam propagation in Goos-H\"anchen and Imbert-Fedorov shifts, Opt. Lett. \textbf{33}, 1437 (2008).

\bibitem{Aiello2009}
A. Aiello, M. Merano, and J. P. Woerdman, Duality between spatial and angular shift in optical reflection, Phys. Rev. A \textbf{80}, 061801(R) (2009).

\bibitem{Qin2011}
Y. Qin, Y. Li, X. Feng, Y.-F. Xiao, H. Yang, and Q. Gong, Observation of the in-plane spin separation of light, Opt. Express \textbf{19}, 9636 (2011).

\bibitem{BARev}
K. Y. Bliokh and A. Aiello, Goos-H\"anchen and Imbert-Fedorov beam shifts: an overview, J. Opt. \textbf{15}, 014001 (2013).



\bibitem{Dennis}
M. R. Dennis and J. B. G\"otte, The analogy between optical beam shifts and quantum weak measurements, New J. Phys. \textbf{14}, 073013 (2012).

\bibitem{Gotte}
J. B. G\"otte and M. R. Dennis, Generalized shifts and weak values for polarization components of reflected light beams, New J. Phys. \textbf{14}, 073016 (2012).

\bibitem{GotteLofflerDennis}
J. B. G\"otte, W. L\"offler, and M. R. Dennis, Eigenpolarizations for giant transverse optical beam shifts, Phys. Rev. Lett. \textbf{112}, 233901 (2014).

\bibitem{AAV}
Y. Aharonov, D. Z. Albert, and L. Vaidman, How the result of a measurement of a component of the spin of a spin-1/2 particle can turn out to be 100, Phys. Rev. Lett. \textbf{60}, 1351 (1988).



\bibitem{DSS}
I. M. Duck, P. M. Stevenson, and E. C. G. Sudarshan, The sense in which a ``weak measurement'' of a spin-1/2 particle's spin component yields a value 100, Phys. Rev. D \textbf{40}, 2112 (1989).

\bibitem{RSH}
N. W. M. Ritchie, J. G. Story, and R. G. Hulet, Realization of a measurement of a ``weak value'', Phys. Rev. Lett. \textbf{66}, 1107 (1991).

\bibitem{Berry435}
M. V. Berry, Lateral and transverse shifts in reflected dipole radiation, Proc. R. Soc. A \textbf{467}, 2500 (2011).

\bibitem{XieSHELinIF}
L. Xie, X. Zhou, X. Qiu, L. Luo, X. Liu, Z. Li, Y. He, J. Du, Z. Zhang, and D. Wang, Unveiling the spin Hall effect of light in Imbert-Fedorov shift at the Brewster angle with weak measurements, Opt. Express \textbf{26}, 22934 (2018).



\bibitem{LiVortexRT}
S. Li, Y. Zhang, Y. Chen, and S. Yu, Reflection and transmission of optical vortex beams at a dielectric interface, in \textit{2013 12th International Conference on Optical Communications and Networks (ICOCN)} (2013) pp. 1--4.

\bibitem{DennisGotteVortex}
M. R. Dennis and J. B. G\"otte, Topological aberration of optical vortex beams: Determining dielectric interfaces by optical singularity shifts, Phys. Rev. Lett. \textbf{109}, 183903 (2012).

\bibitem{YavorskyBrasselet}
M. Yavorsky and E. Brasselet, Polarization and topological charge conversion of exact optical vortex beams at normal incidence on planar dielectric interfaces, Opt. Lett. \textbf{37}, 3810 (2012).

\bibitem{VortexBrewster}
R. Barczyk, S. Nechayev, M. A. Butt, G. Leuchs, and P. Banzer, Vectorial vortex generation and phase singularities upon Brewster reflection, Phys. Rev. A \textbf{99}, 063820 (2019).



\bibitem{LMB}
L. M. Brekhovskikh, \textit{Waves in Layered Media} (Academic Press, NY/London, 1960).

\bibitem{P1956}
S. Pancharatnam, Generalized theory of interference, and it's applications: Part i. coherent pencils, Proc. Ind. Acad. Sci. A \textbf{44}, 247--262 (1956).

\bibitem{Berry}
M. V. Berry, Quantal phase factors accompanying adiabatic changes, Proc. R. Soc. A \textbf{392}, 45 (1984).

\bibitem{Berry1987}
M. V. Berry, The adiabatic phase and Pancharatnam's phase for polarized light, J. Mod. Opt. \textbf{34}, 1401--1407 (1987).

\bibitem{Shapere}
A. Shapere and F. Wilczek, \textit{Geometric Phases in Physics} (World Scientific, Singapore, 1989).

\bibitem{Bliokh2008}
K. Y. Bliokh, Y. Gorodetski, V. Kleiner, and E. Hasman, Coriolis effect in optics: Unified geometric phase and spin-Hall effect, Phys. Rev. Lett. \textbf{101}, 030404 (2008).



\bibitem{Bliokh2009}
K. Y. Bliokh, Geometrodynamics of polarized light: Berry phase and spin Hall effect in a gradient-index medium, J. Opt. A: Pure Appl. Opt. \textbf{11}, 094009 (2009).

\bibitem{BA2010}
K. Y. Bliokh, M. A. Alonso, E. A. Ostrovskaya, and A. Aiello, Angular momenta and spin-orbit interaction of nonparaxial light in free space, Phys. Rev. A \textbf{82}, 063825 (2010).

\bibitem{CLEO2020}
A. Debnath and N. K. Viswanathan, Observation of polarization singularities in a Brewster-reflected paraxial beam, in \textit{Conference on Lasers and Electro-Optics} (Optical Society of America, 2020) p. JTh2E.1.

\bibitem{FDTDBook}
A. Taflove and S. C. Hagness, \textit{Computational Electrodynamics: The Finite-Difference Time-Domain Method} (Artech House, Inc., MA, 2005), 3rd ed.

\bibitem{FOOTNOTE_VirtualSource}
\textbf{Note:} It is to be noticed that the point $O$ acts only as a virtual point source, where the field $\tilde{\mathbf{E}}_{i}^{(I)}$ never physically exists. However, assigning a zero reference phase to the point $O$ is a matter of geometrical convenience without any loss of generality.

\bibitem{FOOTNOTE_AmpDirection}
\textbf{Note:} It is to be noticed that, only identifying a given $\mathcal{E}$ as the amplitude value without referring to the corresponding unit vector direction (e.g. $\hat{\mathbf{x}}$ in Eq. (\ref{Amp_E=E_1}) ) results in a sign-ambiguity; because, in a physical system, one can choose the opposite unit vector (e.g. $\hat{\mathbf{x}}' = -\hat{\mathbf{x}}$ for the case of Eq. (\ref{Amp_E=E_1}) ) as the reference direction, with respect to which the amplitude value is $\mathcal{E}' = -\mathcal{E}$. This sign-ambiguity is the reason behind the sign-difference between the Fresnel $r_{TM}$ expressions in Refs. \cite{Jackson, BornWolf} and Ref. \cite{SalehTeich}. We eliminate this ambiguity by always using vector expressions, as discussed in Section \ref{Sec_Derivation}.

\bibitem{FOOTNOTE_2elem}
\textbf{Note:} The phase term $\left( \mathbf{k} \cdot \Delta\mathbf{r} - \omega t + \Phi_O \right)$ is the same for both the elements, since they are on the same wavefront. Hence, this phase term does not contribute to the calculation of the geometric phase difference between the two concerned elements. This justifies the omission of the phase term $\left( \mathbf{k} \cdot \Delta\mathbf{r} - \omega t + \Phi_O \right)$ for the purpose of the present demonstration.

\bibitem{FOOTNOTE_Superposition}
\textbf{Note:} This virtual superposition is considered only to compare the two fields to calculate their geometric phase difference --- the two surface elements never superpose in the actual physical system. For the Fourier decomposition case, this superposition happens in the actual physical system only for $\delta = 0$.

\end{thebibliography}
\end{document}